\documentclass[%
 reprint,
 amsmath,amssymb,
 aps,
]{revtex4-1}
\usepackage{float}
\usepackage{caption}
\usepackage{subcaption}
\usepackage{xcolor}
\usepackage{graphicx}% Include figure files
\usepackage{dcolumn}% Align table columns on decimal point
\usepackage{bm}% bold math
\DeclareMathOperator\erf{erf}
\DeclareMathOperator\erfc{erfc}
\DeclareMathOperator\erfi{erfi}

\begin{document}

\preprint{APS/123-QED}

\title{Random Process Theory Approach to Geometric Heterogeneous Surfaces:\\ 
	Effective Fluid-Solid Interaction}% Force line breaks with \\

\author{Aleksey Khlyupin}
 \altaffiliation[]{Schlumberger Moscow Research Center; 13, Pudovkina str., Moscow 119285, Russia 
 	\\ 
 	Moscow Institute of Physics and Technology; 9, Institutskiy per., Dolgoprudny, Moscow Region, 141701, Russia}%Lines break automatically or can be forced with \\ 
\author{Timur Aslyamov}%
 \email{t.aslyamov@gmail.com; taslyamov@slb.com}
 \altaffiliation[]{Schlumberger Moscow Research Center; 13, Pudovkina str., Moscow 119285, Russia}
\affiliation{%
Schlumberger Moscow Research Center; 13, Pudovkina str., Moscow 119285, Russia 
\\
Moscow Institute of Physics and Technology; 9, Institutskiy per., Dolgoprudny, Moscow Region, 141701, Russia
}%
\date{\today}% It is always \today, today,

\begin{abstract}
Realistic fluid-solid interaction potentials are essential in description of confined fluids especially in the case of geometric heterogeneous surfaces. Correlated random field is considered as a model of random surface with high geometric roughness. We provide the general theory of effective coarse-grained fluid-solid potential by proper averaging of the free energy of fluid molecules which interact with the solid media. This procedure is largely based on the theory of random processes. We apply first passage time probability problem and assume the local Markov properties of random surfaces. General expression of effective fluid-solid potential is obtained.  In the case of small surface irregularities analytical approximation for effective potential is proposed. Both amorphous materials with large surface roughness and crystalline solids with several types of fcc lattices are considered. It is shown that the wider the lattice spacing in terms of molecular diameter of the fluid, the more obtained potentials differ from classical ones. A comparison with published Monte-Carlo simulations shows good qualitative agreement with the theory predictions. The work provides a promising approach to explore how the random geometric heterogeneity affects on thermodynamic properties of the fluids.

\end{abstract}

\pacs{Valid PACS appear here}% PACS, the Physics and Astronomy
\maketitle

%\tableofcontents

\section{Introduction}

Interaction between fluid and solid molecules is essential for various fundamental phenomena: adsorption, capillary condensation, wetting and etc. In spite of long history of these problems major results were obtained for idealized smooth surfaces of the solid. However, real surfaces are usually rough, so influence of geometry on adsorption and other thermodynamic processes is actively investigated in recent years \cite{garoff1989effects, quere2002rough, beaglehole1994extrinsic, netz1997roughness, kuchta2008melting, coasne2013adsorption, urrutia2014bending, schoen1997structure, urrutia2014fluids, henderson2004statistical, ustinov2006pore, ravikovitch2006density, jagiello2013carbon, khlyupin2016effects, aslyamov2014complex}. 

Wetting properties of fluid near heterogeneous surfaces has been studied in details \cite{garoff1989effects, quere2002rough}. It was shown both experimentally \cite{beaglehole1994extrinsic} and theoretically \cite{netz1997roughness} that the roughness can induce wetting transition at the conditions corresponding to the lack of wetting for smooth surface. Also, experiments with neutron scattering and Monte Carlo simulations show strong influence of the surface roughness on melting process \cite{kuchta2008melting}. 
In work \cite{coasne2013adsorption} the review of intrusion and freezing in porous silica with both simple and disordered porous geometry is presented. %Behavior of fluids confined inside vessels with various form is topic of active research. 
Several works are devoted to fluid confined by both smooth spherical and cylindrical pores \cite{urrutia2014bending} shapes and fluid near abrupt wedges \cite{schoen1997structure, urrutia2014fluids, henderson2004statistical}. %The form of irregularity can be important, behavior of fluids confined in wedges or by edges is actively investigated in recent years.  

Additional motivation to study heterogeneous surface is adsorption in real porous materials. 
%It is well known, that complex structure of real pore materials can be analyzed in terms of pore size distribution. This method has long history, since work Seaton1989 PSD is the solution of the adsorption integral equation, which contains the experimental amount of adsorbed molecules in the sample and theoretical isotherms expressed as the fluid densities in the pores. 
Various methods for theoretical description of fluids inside the pore were developed. However in spite of the variety all standard methods have similar assumption, that solid surface is flat, smooth homogeneous plane. Hence several authors \cite{ustinov2006pore, ravikovitch2006density, jagiello2013carbon} noted, that theoretical adsorption isotherm has multi-step form due to multilayer structure of adsorbed fluid. Then, in the case of pore materials with wide pore size distribution two artifacts appeared: calculated adsorption isotherms have typical deviation from experimental one; obtained pore size distribution has the sharp gap corresponding to $10 \AA$ pores size. Thus theoretical adsorption isotherms obtained for pores with smooth surfaces are unable to fit the experimental data for real materials. In order to avoid assumptions about solid surface in works \cite{ustinov2006pore, ravikovitch2006density} authors considered the carbon material as an amorphous media with variable one-dimensional density near the surface. Therefore the solid density is represented as rapidly decreasing function of the distance to the surface. Thus, the heterogeneity is described by only one parameter corresponding to the value of characteristic roughness. These modifications have acceptable agreement with experimental measurements for nongraphitized carbon.  Jagiello and Olivier \cite{jagiello2013carbon} considered two-dimensional heterogeneous surface, taking into account that carbon structure consists of curved graphene layers. In order to describe lateral corrugation authors defined explicit coordinate function for structure of surface. %However, natural surfaces is not well structured object and contain roughness of various scale. 

With the development of computer methods of research it has become possible to study a random surface geometry of the pore space of three-dimensional models, for example, by methods of fractal analysis \cite{khlyupin2015fractal}. In the classical work \cite{avnir1984molecular} surface area of the porous media was determined by experimental adsorption isotherms, which allows drawing a conclusion about their fractal properties. Now it is appropriate to represent rough surface as random process\cite{savva2010two, herminghaus2012universal, persson2004nature}. The random rough surface is characterized by two parameters corresponding to variance and correlation function of random process. Results of artificial solid generation in such way are very similar with experimentally observed surface profiles. Rough surfaces are commonly divided into deterministic and random surfaces. Deterministic surfaces have a pre-scribed shape, usually of a simple form such as triangular, rectangular or sinusoidal. Random surfaces, are stochastic and are usually characterized using terms from probability theory such as distribution function or the statistical moments. 

Random surfaces approach can be found in other physical problems. For example, the diffusive motion of Brownian particles near irregular interfaces is important in various transport phenomena in nature and industry. Most diffusion-reaction processes in confinied interfacial systems involve a sequence of Brownian flights in the bulk, connected by successive hits with the interface (Brownian bridges). In the work \cite{levitz2006brownian} theoretical and numerical analysis of bridge statistics was presented, and it was shown, that the results is directly related to the stochastic properties of the surface.
Also roughness of the surfaces is highly important in the theory and applications of scattering of electromagnetic waves \cite{beckmann1987scattering}. In the case of light scattering from random rough surfaces the ruling probability density distributions are given by the height distribution functions and the correlation functions describing the rough surfaces and the ray distribution describing the incident light \cite{bergstrom2007ray}. For modelling and simulation purposes random rough surfaces can be generated using a method outlined by Garcia and Stoll \cite{garcia1984monte}, where an uncorrelated distribution of surface points using a random number generator (i.e. white noise) is convolved with a Gaussian filter to achieve correlation.
%Rough surfaces are commonly divided into deterministic and random surfaces. Deterministic surfaces have a pre-scribed shape, usually of a simple form such as triangular, rectangular or sinusoidal that is easy to implement but lack of realism. 

A lot of phenomena of surface physics can be described by effective potential for interaction between fluid molecule and solid media. Indeed, in the case of smooth surface and Lennard Jones (12-6) interaction well known potentials like (10-4), (9-3) are results of integrating over single-layer molecular solid and multilayer ones, correspondingly. Also in the case of multi-layered solid more accurate potential (10-4-3) was obtained by Steel \cite{steele1973physical}. These potentials are widely used in the density functional theory (DFT), calculations of wetting phenomena and disjoining pressure. Calculations of the effective potential for solids with well-defined structure was performed in work \cite{forte2014effective}. According to this work total interaction between fluid molecule and solid media can be replaced by effective potential as function only of the distance from a molecule to the surface. The method of averaging potential is based on the comparison of the Helmholtz free energy obtained by accurate partition function of initial system and one which is result of approximated partition function with effective potential. %This method differs from both configuration averaging $ <U>=N_{conf}^{-1}\sum_{k=1}^{N_{conf}}u_k$, where $N_{conf}$ is the number of all configurations of system, and canonical averaging with the following Gibbs's weight
%\begin{eqnarray}
%<U>=\frac{\sum_{k=1}^{N_{conf}}u_k e^{-\beta u_k}}{\sum_{k=1}^{N_{conf}} e^{-\beta u_k}}
%\end{eqnarray}
The free energy average technique -- is an approach to link the free energy of the system to one of a system with fewer degrees of freedom. Effective potentials of this type can be developed by ensuring that the partition function of the integrated representation corresponds to that of the explicit system. This type of averaging is commonly employed to obtain the reference-system potentials in different %in reference average Mayer-function perturbation theory [6] and in other 
perturbation theories \cite{perram1974perturbation, verlet1974perturbation, steinhauser1981molecular}. %For example attempts to develop a perturbation theory for dense fluids presented in [10]. 
Free-energy average potentials are also widely used to describe effective interactions in colloidal, polymeric and biomolecular systems \cite{voth2008coarse}. %However, the extension of this approach to development the effective fluid-surface interactions is overlooked. Classical potentials like (9-3), (10-4) or (10-4-3) Steel potential are widely used in the DFT modeling of fluids near solid with flat surfaces. 
In the work \cite{forte2014effective} authors demonstrated that free-energy average potential is essential to accurate description of adsorption on solids with definite structure at low and moderate temperature. %A comparison with classical effective potentials was made by adequacy in adsorption description on different solid lattices based on Grand Canonical Monte Carlo simulation. 
Corresponding effective potentials %in the work \cite{forte2014effective} 
were obtained by numerical simulations of Lennard-Jones fluid by accounting for pair interactions of a fluid particle with every particle of the solid arranged in a given ordered lattice.
% and did not assume materials with large heterogeneity. %But even on crystalline faces with single defect Ii was shown that the higher heterogeneity of the surface the less adequate simple classical effective potentials are to describe fluid behavior.
For such cases it was shown by Monte-Carlo simulations that effective free energy averaged potentials are softer than classical ones (10-4, 9-3, 10-4-3 potentials). This reflects the fact that the less corrugated and smooth surface is the more fluid particles are prone to explore the gaps of the surface, especially at low temperatures. Thus even for surfaces with small heterogeneity correct averaged potentials are quite different from ideal smooth case. Therefore in the case of surfaces with large geometric heterogeneity realistic fluid-solid potentials are necessary.
%Realistic fluid-solid potentials are essential in the development of theories of adsorption on surfaces with large geometric heterogeneity and randomness. 

\begin{figure}[H]
	\includegraphics[width=8.5cm]{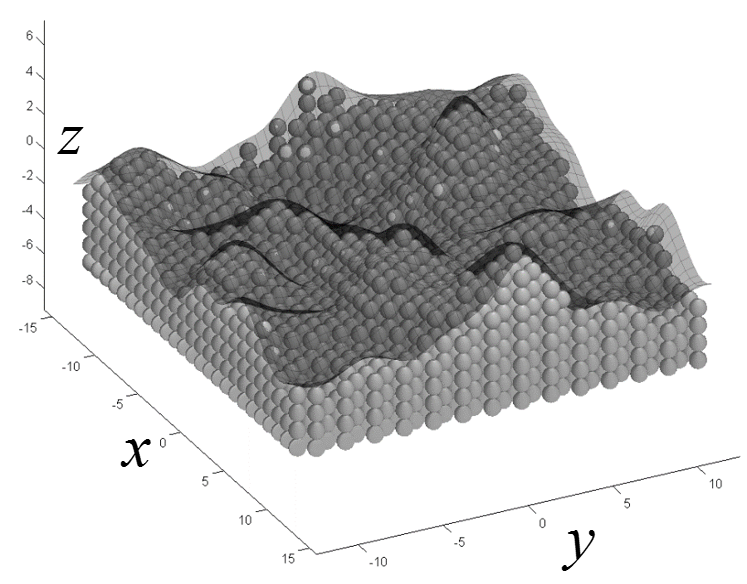}
	\caption{\label{fig_1} Example of solid media with geometric heterogeneous surface. Gray balls illustrate solid molecules. Solid surface can be described by certain realization of the random field. The system of coordinate is presented in reduced units with respect to diameter of fluid particle $d$.}
\end{figure}

In our current manuscript we provide the general theory and calculations of leading-order term of effective coarse-grained fluid-solid potential. The approach based on the mapping from average free energy of fluid molecules near the random high heterogeneous surfaces. In contrast to the case when the particle is far from the solid and one may use only general properties of random process which describe the surface we investigate particles in the vicinity of the surfaces and even their penetration into the solid so both general and local properties should be used. To obtain the average distance between solid and fluid particle inside it we applied first passage time probability theory and assume the local Markov properties of the random surface. Also we use the expansion of the two-point probability density of a random process in a series of the correlation function to obtain the conditional probability of random solid density in each z-coordinate “slice”. It may be applied for wide range of correlation functions of the random solid surface which could be obtained from experiments (for example X-ray measurements). Characteristic illustration of considered solid molecular media and corresponding heterogeneous surface can be found in Fig.~\ref{fig_1}.

In order to examine the effective fluid-solid potentials we provide several calculations in the case of large irregularities of random surface. Also the general formula for potentials was simplified to obtain the approximated solution for the case of small surface irregularities. Effective potentials for solids with several types of face cubic centered lattices (111), (100), (110) were calculated. 
It was shown that the wider is the lattice spacing in terms of molecular diameter of the fluid, than obtained potentials the grater is deviation from classical ones (Steel potential or 9-3 potential \cite{israelachvili2011intermolecular}). Also this effect was demonstrated in \cite{forte2014effective} by fully atomistic Monte-Carlo simulations and the comparison shown good agreement of our theory predictions and simulation. Obtained method provides a promising approach to explore how random geometry heterogeneity affects on thermodynamic properties of the fluid which is highly desirable in any density functional theory calculations.

This paper is arranged as follows: partition function of the considered system and general result of  averaging procedure in terms of functional integral are given in the next section. Then, application of general method is demonstrated as sketch of further calculations and three major steps of calculations are formulated. Thus, section of calculations is divided on these systematic steps. Each obtained exact result of subsections is complimented by exact analytical expression in more convenient form or analytical approximation. As result significant deviation from flat smooth surface was obtained for the cases of both large and small scales of roughness. Furthermore, simplification was done to approximate the effective fluid-solid interaction potential in the case of molecular size roughness. Results were obtained for two groups of surfaces: exact formula is used in the case of large deviations from flat smooth surface; in the case of small scale roughness general result is simplified and approximated analytical expression is obtained for roughness similar with molecular size. 

\section{General theory} 

Let us consider a molecule of fluid, interacting with solid phase. Fluid molecule is a sphere with diameter $d$. The solid is represented by a system of $M$ noninteracting molecules located at the sites of three dimensional lattice. 
In this work pair-interactions between a fluid and solid molecules are considered. Solid media has sufficiently large surface, which corresponds to certain realization of random process $\mathcal{Z}(r)$, where $r^2=x^2+y^2$, system of coordinates is shown in Fig~\ref{fig_1}. 
Without loosing of generality we assume that the mean value of $\mathcal{Z}$ equals zero. 
Thus canonical partition function of this system has the following form:
\begin{eqnarray}
Q=Q_f(T)\int d\vec{r}_f e^{-\beta U(\vec{r}_f, \vec{r}_{s}^{(1)},...,\vec{r}_{s}^{(M)})} 
\end{eqnarray}
where $\beta=1/k_B T$, $T$ is the temperature, $k_B$ is  Boltzmann constant,  $Q(f)$ is the factor concluding thermal fluctuations of a fluid molecule, below for simplicity this factor will be left out; $U$ is the total potential of interactions between fluid molecule at point $\vec{r}_f=(x_f, y_f, z_f)$ and solid molecules at fixed points 
$\vec{r}_{s}^{(1)},...,\vec{r}_{s}^{(M)}$. The total potential can be represented as sum of pairwise additive interactions $U=\sum_{i=1}^{M}u_{fs}(\vec{r}_f,\vec{r}_s^{(i)})$. 
 %Let us put z-coordinate at normal direction to solid media surface and 
Let us consider a fluid particle lying at fixed z-coordinate, then corresponding partition function is
\begin{eqnarray}
\label{PartFunction_2}
\iint\limits_{\Omega(z)}dx_fdy_f \exp\left[-\beta \sum_{i=1}^{M}u_{fs}(\vec{r}_f,\vec{r}_s^{(i)})\right]
\end{eqnarray}
where $\Omega(z)$ is the configuration space of fluid molecule restricted by certain realization of random process $\mathcal{Z}(r)$ at fixed $z$-coordinate.  
More compact form for integration can be introduced $\iint_{\Omega(z)}dx_fdy_f ...= \int d\Omega...$ . One can expand exponential term in \eqref{PartFunction_2}, then expression \eqref{PartFunction_2} in the continuum limit has the following form:
\begin{eqnarray}
\label{PartFunction_3}
&Q(z, \mathcal{Z}(r))=\int d\Omega\left[1-\beta\int d\vec{r}_su_{fs}(\vec{r}_f, \vec{r}_s)\rho(r_s, Z(r))+ \nonumber \right. \\
&\left. +O(\beta^2)\right]
\end{eqnarray}
where $\rho(\vec{r}_s, \mathcal{Z})$ is the solid density, the number of solid molecules at volume $d\vec{r}_s$. It is important to note, that this function also depends on random process $ \mathcal{Z}(r)$. Thus, the Helmholtz free energy corresponding to certain realization of random process $ \mathcal{Z}$ can be found from the following expression:
\begin{eqnarray}
\label{F_def}
\beta F(z,\mathcal{Z}(r))=-\ln Q(z;\mathcal{Z}(r))
\end{eqnarray}
Thermodynamic properties of this system are defined from free energy 
averaged over all realizations of random geometry:
\begin{eqnarray}
\label{Aver_F_def}
\beta \left\langle F(z)\right\rangle_\mathcal{Z}\equiv-\int...\int\ln Q(z;\mathcal{Z}(r))P(\mathcal{Z})\prod_r d\mathcal{Z}(r) 
\end{eqnarray}
where integrals imply functional integration over variations of $\mathcal{Z}$ at each point $r$. $P(\mathcal{Z})$ is probability, that certain $\mathcal{Z}$ takes place.
One can expand logarithm in expression \eqref{Aver_F_def} taking into account \eqref{PartFunction_3}. Thus averaged free energy can be written as:
\begin{eqnarray}
\label{AverFreeEnergy}
&\beta \left\langle F\left(z\right)\right\rangle_\mathcal{Z}=-\left\langle \ln \parallel\Omega(z,\mathcal{Z})\parallel\right\rangle_\mathcal{Z}+\beta \left\langle \parallel\Omega(z,\mathcal{Z})\parallel^{-1} \times \right. \nonumber \\
&\left.\times \int d\Omega\int d\vec{r}_s u_{fs}(\vec{r}_f,\vec{r}_s)\rho\left(\vec{r}_s, \mathcal{Z}\right)\right\rangle_\mathcal{Z} +O(\beta^2)
\end{eqnarray}
where $\parallel\Omega(z,\mathcal{Z})\parallel=\int d\Omega$ is the total area of fluid's configuration space at level with fixed $z$ for certain realization of $\mathcal{Z}$.

On the other hand it is possible to substitute total pairwise intermolecular interaction by external field, which implies desired effective fluid-solid potential. 
Let us consider fluid particle with such external field, which depends on z-coordinate $U^{eff}_{fs}(z)$. This system is described by the same random geometry. Thus partition function for fixed $z$ can be written as
\begin{eqnarray}
Q^{eff}(z)=\int d\Omega e^{-\beta U^{eff}_{fs}(z)}
\end{eqnarray}
After simple calculations, expression for averaged free energy takes the form:
\begin{align}
\label{AverFreeEnergyEffField}
\beta \left\langle F^{eff}\left(z\right)\right\rangle_{\mathcal{Z}}=-\left\langle \ln \parallel\Omega(z,\mathcal{Z})\parallel\right\rangle_\mathcal{Z} +\beta  U^{eff}_{fs}(z)
\end{align}

Expressions of exact \eqref{AverFreeEnergy} and simplified \eqref{AverFreeEnergyEffField} averaged energies can be equated $\left\langle\beta F^{eff}\left(z\right)\right\rangle_Z=\left\langle\beta F\left(z\right)\right\rangle_Z$, thus effective potential of fluif-solid interaction has the following form
\begin{eqnarray}
\label{EffPotential_1}
&U^{eff}_{fs}(z)=\left\langle \dfrac{\int d\Omega\int d\vec{r}_s u_{fs}(\vec{r}_f,\vec{r}_s)\rho\left(\vec{r}_s;\mathcal{Z}\right)}{\parallel\Omega(z;\mathcal{Z})\parallel}\right\rangle_\mathcal{Z} +O(\beta)\nonumber \\
\end{eqnarray}

The integration region for fluid molecule in expression \eqref{EffPotential_1} is union of non-crossing domains $\Omega_i$ with random sizes, $\Omega(z)=\bigcup_{i=1}^{\infty}\Omega_i(z)$. Permitted random regions $\Omega_i$ are induced by random binary field which is a slice of random process $\mathcal{Z}$ at level $z$. One can rewrite integral in \eqref{EffPotential_1} taking into account structure of $\Omega$
\begin{eqnarray}
&\int d\Omega \int d\vec{r}_s u_{fs}(\vec{r}_f,\vec{r}_s)\rho\left(\vec{r}_s;\mathcal{Z}\right)= \nonumber \\
&=\sum_{i=1}^{\infty}\int d\Omega_i \int d\vec{r}_s u_{fs}(\vec{r}_f,\vec{r}_s)\rho\left(\vec{r}_s;\mathcal{Z}\right)= \\
&=\sum_{i=1}^{\infty} \parallel\Omega_i(z)\parallel U(\Omega_i(z)) \nonumber
\end{eqnarray}
where $U(\Omega_i(z))$ is average potential at domain $\Omega_i(z)$. It is possible to make simplification similar to mean field approximation, then
\begin{eqnarray}
&\sum_{i=1}^{\infty} \parallel\Omega_i(z)\parallel U(\Omega_i(z)) \simeq \sum_{i=1}^{\infty} \parallel\Omega_i(z)\parallel U(\bar{\Omega}(z))= \nonumber \\
&=U(\bar{\Omega}(z))\parallel\Omega(z)\parallel
\end{eqnarray} 
where $\bar{\Omega}(z)$ is domain with the average (characteristic) size. This average size depends on the properties of random process and the coordinate $z$. Thus, effective potential of interaction between fluid molecule and solid with random heterogeneous surface has the following form:
\begin{eqnarray}
\label{EffPotential_2}
&U_{fs}^{eff}(z)=\int dr_s u_{fs}(r_f, r_s) \times \\ \nonumber
&\times \int d\rho(r_s)\rho(r_s)P\left(\rho(r_s)|r_f\in\bar{\Omega}(z)\right)
\end{eqnarray}
where $P\left(\rho(r_s)|r_f\in\bar{\Omega}(z)\right)$ is probability density of $\rho(r_s)$ under condition that fluid particle lies at characteristic domain $\bar{\Omega}(z)$. Thus, effective fluid-solid potential reflects the random surface properties by probability density $P$ and average size of $\bar{\Omega}$.

\section{\label{Strategy} Strategy of calculations}

In this section we provide a scheme of step by step calculations.
To simplify further calculation one dimensional random process $\mathcal{Z}(x)$ is considered. Then a slice at any level $h$ is random binary field $\xi_h(x)=\theta\left(Z(x)-h\right)$. 
\begin{equation}
\xi_h(x)=\begin{cases}
1, Z(x) \geq h\\
0, Z(x)< h
\end{cases} \nonumber
\end{equation}

\begin{figure}[H]
	\includegraphics[width=8.5cm]{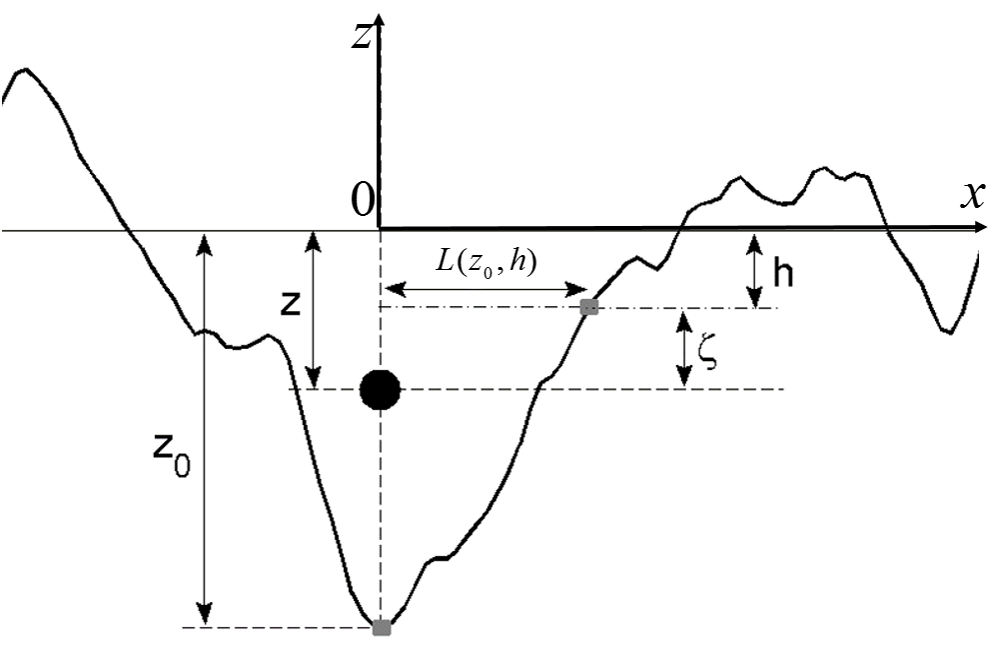}
	\caption{\label{fig_2} Illustration of the molecule (bold dot) and some realization of random process corresponding to heterogeneous surface (solid curve). In this case the slice at level $h$ is higher than fluid particle's one. $L(z_0,h)$ is the minimal length along an axis $x$ such that    $\mathcal{Z}(L)=h$.}
\end{figure}

Let us consider a fluid particle at level $z$. The origin of x-axis coincides with the molecule position and this molecule is below the slice level $h$ (see Fig.~\ref{fig_2} for explanation). 

Thus, contribution to the total potential of fluid-solid interaction from the slice of the width $dh$ is:
\begin{align}
\label{Ihz}
I(h,z)=\rho_s dh \int_{-\infty}^{\infty}\xi_h(x)u_{fs}(r(x))dx
\end{align}

where $u_{fs}(r)$ is pair-intermolecular potential, $r=\sqrt{(z-h)^2+x^2}$ is the distance between considered molecule and certain point of solid media. Expression \eqref{Ihz} corresponds to some realization of random process, so averaging over distribution of $\xi_h(x)$ is needed. For this reason one can use information about local structure of binary field $\xi_h(x)$ near $x=0$. Since in considered case $\mathcal{Z}(0)=z_0<z<h$ then $\xi_h(0)=0$. For fixed position of fluid particle and some realization of random process  let us define length $L(z_0,h)$ as $L=\{x: \xi_h(x)=0 \,\, \text{for} \,\, \forall x\in[0,L]\}$ . Simple illustration can be found in Fig.~\ref{fig_2}.  
Thus, the region inside $(0,L)$-interval can be excluded from integration over layer $h$\eqref{Ihz}:
\begin{align}
\label{Ihz2}
I(h,z)=2\rho_s dh \int_{L}^{\infty}\xi_h(x)u_{fs}(r(x))dx
\end{align}

In order to calculate effective potential according to \eqref{EffPotential_2} average size of $\bar{\Omega}(z)$ and probability density $P[\rho(r_s)|r_f\in\bar{\Omega}(z)]$ are needed.
It is known that $\xi_h(L)=1$, then averaging of \eqref{Ihz2} has to be executed over probability
$P_{11}(h,s)\equiv P(\xi_h(s)=1|\xi_h(0)=1)$, which is probability that $\xi_h(s)=1$ under condition $\xi_h(0)=1$. The average size of $\bar{\Omega}(z)$ is defined by length $\mathcal{L}(z,h)$ which is averaged $L(z_0, h)$ taking into account correlation properties of the process $\mathcal{Z}(x)$. 
Thus, averaged expression of \eqref{Ihz2} takes the form:
\begin{align}
\label{aver_Ihz}
\left\langle I(h,z)\right\rangle _{\xi}=2\rho_s dh\int_{\mathcal{L}(z,h)}^{\infty}dxu_{fs}(r(x))P_{11}(h,x-\mathcal{L}(z,h))
\end{align}

Now it is possible to formulate the sketch of calculation of the total interaction potential between a molecule and solid media with heterogeneous surface
\begin{itemize}
	\item calculate functions $P_{11}(h,s)$ and $\mathcal{L}(z,h)$ for each layer at level $h$
	\item perform averaging over probability function $P_{11}(h, s)$ for certain layer at level $h$
	\item final step is the integration over solid media
\end{itemize}

\section{Calculations}
%Here and below the case of 2D radial symmetry surface is considered. Accordingly above section the first step is calculation of condition probability:

\subsection{\label{Probability} Conditional probability} 
In this part of work conditional probability $P_{11}(h, s)$ is derived. One can note that condition probability $P_{11}(h, s)$ is equal to correlation function for random field $\xi_h(x)$:
\begin{eqnarray}
P_{11}(h, s)\equiv P(\xi_h(s)=1|\xi_h(0)=1)=\left\langle \xi_h(0)\xi_h(s)\right\rangle \nonumber
\end{eqnarray}

Let's consider random binary field $\xi_h(x)$ as a slice of random process $\mathcal{Z}(x)$ at level h:
\begin{eqnarray}
\xi_h(x)=\begin{cases}
1, Z(x) \geq h\\
0, Z(x)< h
\end{cases}
\end{eqnarray} 
Then, condition probability $P_{11}(h,s)$ can be written as the following double integral:
 \begin{eqnarray}
 \label{P11_def}
 &P_{11}(h, s)\equiv P(Z(0)\geq h|Z(s)\geq h)= \\
 &=\Xi^{-1} \iint_{-\infty}^{\infty}F(z_1)F(z_2)w_z^{(2)}(z_1, x;z_2,x+s)dz_1dz_2 \nonumber 
  \end{eqnarray}
  \begin{eqnarray}
   &\Xi=\iint_{-\infty}^{\infty}w_z^{(2)}(z_1, x ;z_2, x+s)dz_1dz_2 
  \end{eqnarray}
where normalization  $\Xi$ and new function $F(x)=1-\theta(x-h)$ were introduced, $w_z^{(2)}$ is two-dimensional density distribution function of the process $\mathcal{Z}(x)$. In the current research stationary Gaussian process is considered. Thus at this case two-dimensional density distribution function with arbitrary correlation function $K(s)$ has the following form
\begin{eqnarray}
\label{Gauss_distr}
w_z^{(2)}(z_1, x ;z_2, x+s)=\dfrac{1}{2\pi\sigma\sqrt{1-K(s)^2}}\times \nonumber \\
\times\exp\left[-\dfrac{z_1^2+z_2^2-2K(s) z_1z_2}{2\sigma^2(1-K(s)^2)}\right]
\end{eqnarray}
Such type of correlation functions can be calculated expanding $\omega_z^{(2)}$ in the orthogonal system series. Thus, the expression \eqref{Gauss_distr} could be rewritten as:
\begin{eqnarray}
\label{Gauss_distr_1}
&w_z^{(2)}(z_1,z_2,K(x))=w_z^{(1)}(z_1)w_z^{(1)}(z_2)\times \nonumber \\
&\times\sum_{n=0}^{\infty}\dfrac{1}{n!}K^{n}(x)H_n(z_1/\sigma)H_n(z_2/\sigma)
\end{eqnarray}
where $H_n$ is n-th order Hermitian polynomial \cite{bateman1955higher}, $w_z^{(1)}$ is one-dimensional Gaussian density
\begin{eqnarray}
\label{Gauss_dens}
w_z^{(1)}(x)=\dfrac{1}{\sqrt{2\pi\sigma^2}}\exp\left(-\dfrac{x^2}{2\sigma^2}\right)
\end{eqnarray}
Taking into account expression \eqref{Gauss_distr_1} one can derive the numerator of \eqref{P11_def}:
\begin{eqnarray}
&\iint_{-\infty}^{\infty}F(z_1)F(z_2)w_z^{(2)}(z_1, x; z_2, x+s)dz_1dz_2= \nonumber \\
&=\sum_{n=0}^{\infty}C_n^2K^n(s)
\end{eqnarray}
where coefficients $C_n$ correspond to the following expression:
\begin{eqnarray}
\label{Cn}
C_n=\sqrt{\dfrac{1}{2^n\pi n!}}H_{n-1}\left(\dfrac{h}{\sigma}\right)\exp\left(-\dfrac{ h^2}{\sigma^2}\right)
\end{eqnarray}
It is easy to note that from expression \eqref{Cn} the value of the normalization $\Xi=C_0$. Taking into account that $K(s)\to 0$ when $s\to\infty$, condition probability $P_{11}(h, s)$ in the limit $s\to\infty$ tends to the value $C_0$ which is average density of binary field $\xi_h(x)$. Thus, for calculation of \eqref{P11_def} one can use the following expression:
\begin{eqnarray}
\label{Probability11}
&P_{11}\left(h, s\right)=C_0+\dfrac{1}{ C_0}e^{-\frac{2 h^2}{\sigma^2}}\sum_{n=1}^{\infty}\dfrac{1}{2^n\pi n!}H_{n-1}^2\left(\frac{ h}{\sigma}\right)K\left(s\right)^n \nonumber 
\\
&C_0=\dfrac{1}{2} \erfc \dfrac{h}{\sqrt{2}\sigma}
\end{eqnarray}
Detailed calculations of sum \eqref{Probability11} can be found in App.~\ref{A1}. Here, let us write the final analytical expression for condition probability:
\begin{align}
\label{P11_result}
P_{11}\left(h, x\right)=C_0-\dfrac{2}{C_0}\left[T\left(\dfrac{\sqrt{2}h}{\sigma}, y(x)\right)-T\left(\dfrac{\sqrt{2}h}{\sigma}, 1\right)\right]
\end{align}
where $T(a, b)$ is special Owen's function \cite{owen1956tables} and implicit dependence on coordinate $x$ is contained inside the following variable 
\begin{eqnarray}
\label{yK}
y=\tan\dfrac{\arccos K(x)}{2}=\sqrt{\dfrac{1-K(x)}{1+K(x)}}
\end{eqnarray}
Analytical expression \eqref{P11_result} with \eqref{yK} is obtained for arbitrary correlation function $K(x)$. For further analysis and calculations the following expression $K(x)=e^{-\alpha x}$ is considered, where $\alpha=\dfrac{1}{\tau}$ is inverse correlation length $\tau$, which depends on type of modeling surface. There are some motivation points for this choice:
\begin{itemize}
	\item In the case when correlation function is unknown. Exponential function is convenient for the calculations and demonstrates the general properties. If there is experimental data for correlation function $K(x)$ then it is possible to approximate obtained data by exponential function with $\tau=K(0)^{-1}\int K(x)dx$.
	\item It is well known in the theory of random processes that the correlation function of stationary Markovian process can be only in form of exponential function \cite{rytov1988principles}.   
\end{itemize} 
\begin{figure}[H]
	\includegraphics[width=8cm]{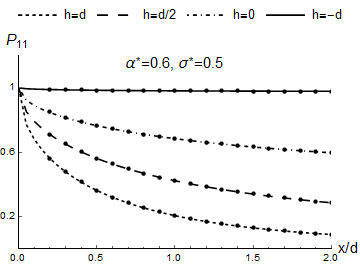}
	\caption{\label{fig_3} Dashed and solid lines are analytical results \eqref{P11_result} of conditional probability $P_{11}$ as function of reduced coordinate $x/d$ calculated for different layers $h$ with reduced parameters $\alpha^*=\alpha d= 0.6, \sigma^*=\sigma/d=0.5$, where $d$ is diameter of fluid particle. Dots correspond to numerical calculations according to \eqref{P11_def}}
\end{figure}

Fig.~\ref{fig_3} illustrates typical behavior of $P_{11}$ as function of reduced coordinate $x/d$. Result \eqref{P11_result} exactly coincides with numerical calculations according to  \eqref{P11_def}. All curves start from the same value $P_{11}(0,h)=1$. 
\begin{figure}[H]
	\includegraphics[width=8cm]{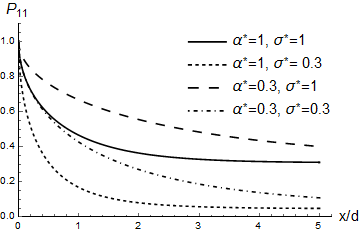}
	\caption{\label{P11_r} Dashed and solid lines are analytical result \eqref{P11_result} of conditional density $P_{11}$ as function of reduced coordinate $x/d$
	calculated for layer $h=0.5d$ with different reduced parameters  $\alpha^*=\alpha d, \sigma^*=\sigma/d$.}
\end{figure}
In Fig.~\ref{P11_r} one can see, that the decline of $P_{11}$ curve depends on  both $\sigma$ and $\alpha$, however, in the limit of large $x$ the curve of $P_{11}$ tends to constant $C_0$ value determined only by $h$ and $\sigma$.

\subsection{\label{Average} Average length}

%Let's consider the particle at point $(0,z)$ and the solid surface as some realization of random process $Z(x)$ Fig.~\ref{Schem_process}, then vertical distance to solid surface is $z-z_0$, Also, how one can see from Fig.~\ref{Schem_process}, the horizontal distance at level $h$ from the particle's vertical to the solid boundary is $L(z_0,h;z)$.

According to general scheme in Sect.~\ref{Strategy}, the next step is calculation of averaged length $\mathcal{L}(h)\equiv \mathcal{L}(z,h)$, which is the average of the length $L(z_0, h)$ over realizations of random process $\mathcal{Z}$. The value of $L(z_0, h)$ implies, that the  binary field $\xi_h(x)$ for $\forall x\in[0,L(h)]$ equals zero $\xi_h(x)=0$, under condition, that the particle is located at point $(0,z)$. That is similar to the known problem of first passage time probability distribution in the theory of random process. In the current problem the distance $L$ plays the role of the time. Thus the calculations consist of two steps: the first aim is the conditional average length $L_{av}(z_0, h)$ taking into account that $\mathcal{Z}(0)=z_0$; the second one is integration of $L_{av}(z_0, h)$ under probability distribution of $z_0$, which results in desired function $\mathcal{L}(z,h)$.

\begin{figure}[H]
	\includegraphics[width=8.5cm]{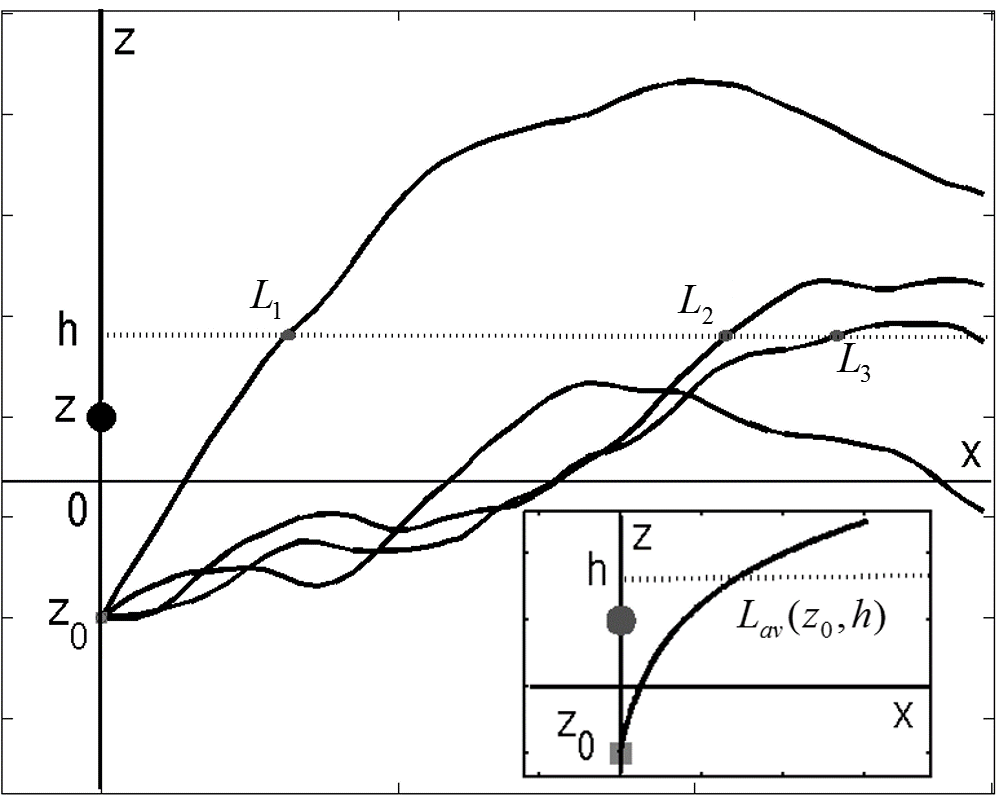}
	\caption{\label{Step1} Illustration of the first step of the averaging procedure. Major figure contains examples of random process realizations with common initial point $z_0$ (solid curves). Black dots on the curves correspond to the first crossing of the boundary at level $h$ by random process. Symbols $L_1, L_2, L_3$ are x-coordinates of these points and correspond to values of $L(z_0, h)$ for illustrated realizations of random process. Inset figure is a sketch of $L_{av}(z_0,h)$ which is average of $L$ over random process realizations with fixed $z_0$}
\end{figure}

\paragraph{The first step} The problem in general form is the finding of distribution function for the length $L(z_0,h)$ when the boundary level $h$ is reached by $\mathcal{Z}$ at the the first "time" under condition, that $\mathcal{Z}(0)=z_0$, where $z_0\leq z \leq h$. Illustration of the first step can be found in Fig.~\ref{Step1}. Exact expression for conditional averaged length $L_{av}$ can be obtained only for the Markovian random process. 
%the random process at first time achieved a certain boundary $h$ in condition that at $z=0$ the value $Z(0)=z_0$, where $z_0\geq z \geq h$.

The theory of Markovian process with boundaries has long history \cite{rice1945mathematical, darling1953first, siegert1951first}. Statistical properties of stationary Markovian process is defined by function $P(z,x|z_0)dz$, which is the probability, that $z_0\leq \mathcal{Z}(x)\leq z$ if $\mathcal{Z}(0)=z_0$. Also, let's introduce the function $F(L, h|z_0)$ -- the probability that the length when the boundary is reached by random process at the first "time" lies at interval $(L, L+dL)$ under condition $\mathcal{Z}(0)=z_0\leq h$. 

There are two general ways to dealing with the problem of calculation the probability distribution $F(L,h|z_0)$ and it’s moments which we are interested in. The first approach may be obtained by observing that the probability
\begin{eqnarray}
\phi(L,h|z_0)=\int_{-\infty}^{h}P(z,x|z_0)dz
\end{eqnarray}
that the random process is at length $L$ in the interval $(-\infty,h)$ is the sum of the probability $f(L,h|z_0)$ that the process did not reach level $h$ at any distance between $(0,L)$, and the probability that it passed $h$ for the first “time” at some length $(0<s<L)$ but returned. Using the relation between cumulative distribution function and probability density function 
\begin{eqnarray}
F(L,h|z_0)=-\dfrac{\partial f(L,h|z_0)}{\partial L}
\end{eqnarray}
it  may be shown that $f(L,h |z_0)$ satisfies the adjoint Fokker-Plank equation (also known as the Backward Kolmogorov Equation BKE). Derivation of this equation in details is presented in \cite{siegert1951first, rytov1988principles}.  Differential equations on the moments of probability density  $F(L,h|z_0 )$ then may be obtained based on this approach as shown in \cite{rytov1988principles}. 
In the present work we will use more simple method to calculate desired moments based on \cite{siegert1951first}. In the case when probability density function of the random process 
$P(z,x|z_0)$ satisfies Fokker-Plank equation (that holds in assumption of Markovian process) the recursion integral formulas for the moments of $F(L,h|z_0)$ may be obtained in the following way. At first, main integral relation may be obtained by classifying random processes $\mathcal{Z}(x)$ for which $\mathcal{Z}(0)=z_0$ according to the distance $s>0$ at which they reach the value $h$ for the first time

\begin{eqnarray}
\label{P_int}
P(z,x|z_0)=\int_{0}^{x}F(s,h|z_0)P(z,x-s|h)ds
\end{eqnarray}
For further analysis, the Laplace transform is needed. The Laplace transform of $F(s,h|z_0)$ can be represented as series of the moments $\left\langle L^n(h|z_0) \right\rangle $
\begin{eqnarray}
\label{Moments}
&F^*(\lambda,h|z_0)=\int_{0}^{\infty}F(s,h|z_0)dL= \nonumber \\
&=\sum_{n=0}^{\infty}\dfrac{(-\lambda)^n}{n!}\left\langle L^n(h|z_0) \right\rangle
\end{eqnarray}
where $f^{\star}$ implies Laplace transform of function $f$. Thus, after Laplace transform of the both parts of \eqref{P_int}, taking into account \eqref{Moments}, expression \eqref{P_int} becomes:
\begin{eqnarray}
\sum_{n=0}^{\infty}\dfrac{(-\lambda)^n}{n!}\left\langle L^n(h|z_0) \right\rangle=\dfrac{P^*(z,\lambda|z_0)}{P^*(z,\lambda|h)}
\end{eqnarray}
Later analysis in terms of Fokker-Plank coefficients is used. If the following limits are existed
\begin{eqnarray}
A(z)=\lim\limits_{\Delta  x\to 0}\dfrac{1}{\Delta x}\int_{-\infty}^{\infty}dy(y-z)P(y,\Delta x|z) \nonumber \\
B(z)=\lim\limits_{\Delta  x\to 0}\dfrac{1}{\Delta x}\int_{-\infty}^{\infty}dy(y-z)^2P(y,\Delta x|z)
\end{eqnarray}
then there are the following equations for probability $P(z,x|z_0)$:
\begin{eqnarray}
&\dfrac{\partial P}{\partial x}=-\dfrac{\partial}{\partial z}\left[A(z)P\right]+\dfrac{1}{2}\dfrac{\partial^2}{\partial z^2}\left[B(z)P\right]  \\
\label{BKE}
&\dfrac{\partial P}{\partial x}=A(z_0)\dfrac{\partial P}{\partial z_0}+\dfrac{1}{2}B(z_0)\dfrac{\partial^2 P}{\partial z_0^2}
\end{eqnarray}
with initial condition $P(z,0|z_0)=\delta(z-z_0)$, and boundary conditions $P(\pm\infty,x|z_0)=0$ and $P(z,x|\pm\infty)=0$ in cases of finite $z_0$ and $z$ respectively. One can rewrite equation \eqref{BKE} as:
\begin{eqnarray}
&\dfrac{\partial P(z,x|z_0)}{\partial x}=\dfrac{1}{2 W(z_0)}\dfrac{\partial}{\partial z_0}\left\lbrace B(z_0)W(z_0)\dfrac{\partial P(z,x|z_0)}{\partial z_0}\right\rbrace \nonumber \\
\end{eqnarray}
were the stationary function was introduced
\begin{eqnarray}
W(s)=\dfrac{\textmd{Const}}{B(s)}\exp\left[\int_{-\infty}^{s}\dfrac{2A(s')}{B(s')}ds'\right]
\end{eqnarray} 
Taking into account the asymptotic condition $B(z_0)W(z_0)\dfrac{\partial P(z,x|z_0)}{\partial z_0}\to 0$ for $z_0\to-\infty$ one can obtained the following expression for $z>z_0$:
\begin{eqnarray}
\label{BKE_1}
&P(z,x|z_0)-P(z,x|h)=\nonumber \\
&=\int_{h}^{z_0}\dfrac{2ds}{B(s)W(s)}\int_{-\infty}^{s}W(t)\dfrac{\partial P(z,x|t)}{\partial x}dt
\end{eqnarray} 
After Laplace transform of the both parts of equation \eqref{BKE_1}, the following expression can be written:
 \begin{eqnarray}
 \label{BKE_2}
 &P(z,x|z_0)^{*}-P(z,x|h)^{*}=\nonumber \\
 &=\lambda\int_{h}^{z_0}\dfrac{2ds}{B(s)W(s)}\int_{-\infty}^{s}W(t)P^{*}(z,\lambda|t)dt
 \end{eqnarray}
Using expression \eqref{Moments} and the fact, that $\left\langle L^0(h|z_0)\right\rangle=1$ on can get:
\begin{eqnarray}
&\sum_{n=0}^{\infty}\dfrac{(-\lambda)^n}{n!}\int_{h}^{z_0}\dfrac{2ds}{B(s)W(s)}\int_{-\infty}^{s}W(t)\left\langle L^{n}(h|t)\right\rangle dt=\nonumber \\
&=\dfrac{1}{\lambda}\left[1-\dfrac{P^*(z,\lambda|z_0)}{P^*(z,\lambda|h)}\right]=-\dfrac{1}{\lambda}\sum_{n=1}^{\infty}\dfrac{(-\lambda)^n}{n!}\left\langle L^n(h|z_0) \right\rangle \nonumber \\
\end{eqnarray}
Thus, the recursion relation for the moments of density distribution of $F(s,h|z_0)$ is obtained:
\begin{eqnarray}
\label{Moments_recursion}
\left\langle L^n(h|z_0) \right\rangle=n \int_{z_0}^{h}\dfrac{2ds}{B(s)W(s)}\int_{-\infty}^{s}W(t)\left\langle L^{n-1}(h|t)\right\rangle dt \nonumber \\
\end{eqnarray}
The average length $L_{av}$ corresponds to the case of $n=1$ in the relation \eqref{Moments_recursion}:
\begin{eqnarray}
\label{_L_av_A_B}
&L_{av}(z_0,h)\equiv\left\langle L^1(h|z_0)\right\rangle= \nonumber \\
&= \int_{z_0}^{h}\dfrac{2ds}{B(s)W(s)}\int_{-\infty}^{s}W(t) dt
\end{eqnarray}
If correlation function $K(x)=e^{-\alpha x}$, it is possible to write simple expressions for Fokker-Plank coefficients: $A(z)=-\alpha z$, $B=2\sigma^2 \alpha$. Thus, the average length has the form:
%Let's consider the nearest to particle $(0,z)$ right (left) boundary as a realization of random process with fixed start point at level $z_0<z$ and final destination in the first point of contact with line $h$. Hence, this process has boundaries in point $h$ and we need calculate average length of reaching this boundaries $L\left(z_0,h\right)$. 
%Without losing generality we can consider Markovian random process. In this case there is exact expression for $L\left(z_0,h\right)$:
\begin{align}
\label{Average_length}
L_{av}\left(z_0,h\right)=\frac{1}{\alpha \sigma^2}\int_{z_0}^{h}e^{\frac{\xi^2}{2\sigma^2}}d\xi\int_{-\infty}^{\xi}d\eta e^{-\frac{\eta^2}{2\sigma^2}}
\end{align} 

Expression \eqref{Average_length} is not convenient for further calculations, because the limits of integration contain variables $z_0, h$. For this reason in Appendix~\ref{Simplification} some simplifications of \eqref{Average_length} were produced, and more appropriate expression can be written as 
\begin{eqnarray}
\label{Average_length_Analytic}
\alpha &L_{av}\left(z_0,h\right)=\int\limits_{0}^{1}d\tau\dfrac{1}{1-\tau^2}\left[e^{\frac{h^2}{2\sigma^2}(1-\tau^2)}-e^{\frac{z_0^2}{2\sigma^2}(1-\tau^2)}\right]+ \nonumber \\
&+\dfrac{\pi}{2}\left[\erfi\dfrac{h^2}{\sqrt{2}\sigma^2}-\erfi\dfrac{z_0^2}{\sqrt{2}\sigma^2}\right]
\end{eqnarray}

\begin{figure}[H]
	\includegraphics[width=8cm]{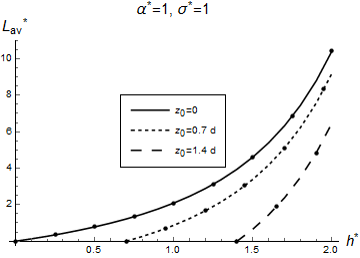}
	\caption{\label{fig_L_z0} Dimensionless average length $L_{av}^*=L_{av}/d$ calculated from expression \eqref{Average_length_Analytic} as function of dimensionless $h^*=h/d$ for different initial points  $z_0$. Dots correspond to numerical calculations according to expression \eqref{Average_length}.}
\end{figure}

Fig.~\ref{fig_L_z0} shows how average length depends on the choice of point $z_0$. As one can see the deeper initial point $z_0$ is, the smaller width of space free from solid media at the same level $h$ would be. Also the figure demonstrates excellent accuracy of simplified expression \eqref{Average_length_Analytic} in comparison with initial exact one \eqref{Average_length}

\begin{figure}[H]
	\includegraphics[width=8.5cm]{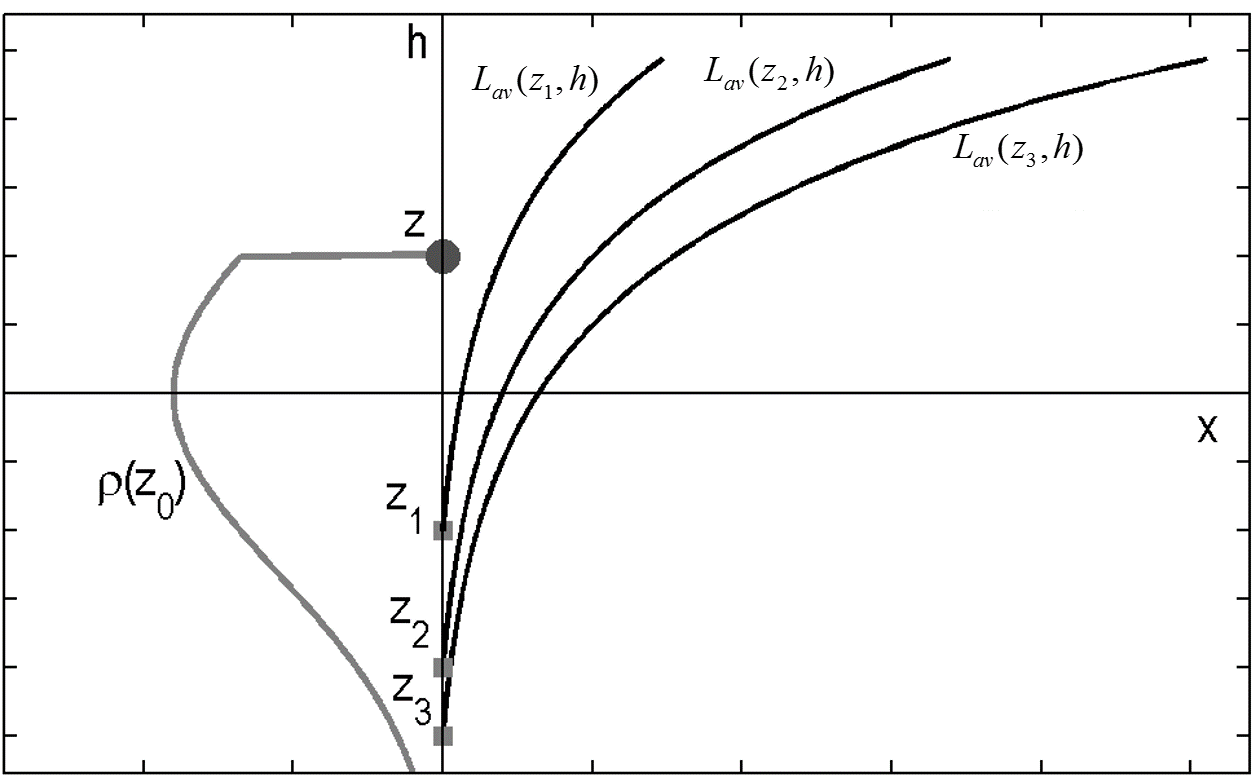}
	\caption{\label{Step2} Illustration of the second step of the averaging. The right side curves correspond to examples of $L_{av}(h,z_0)$ calculated for different started points: $z_1, z_2, z_3$. On the left side of the figure, gray curve is sketch of probability distribution $\rho(z_0)$ of the initial point, $\rho(z_0>z)\equiv0$.   }
\end{figure}

\paragraph{The second step} In fact initial point $z_0$ is unknown except condition $z_0<z$. For this reason the next step is integration of $L_{av}(z_0,h)$ over $z_0$ under condition probability distribution $P(z_0|z_0<z)$ defined by the model of random process $\mathcal{Z}(x)$. Demonstration of the second step can be found in Fig.~\ref{Step2}. Important to note that averaging procedure strongly depends on relative positions of fluid particle $z$ and level $h$
\begin{eqnarray}
\mathcal{L}(z, h)=\begin{cases}
\mathcal{L}^-(z, h), z \leq h\\
\mathcal{L}^+(z, h), z > h
\end{cases} \nonumber
\end{eqnarray} 
Then, in the case $z\leq h$:
\begin{eqnarray}
&\mathcal{L}^-(z,h)\equiv\int L_{av}(z_0, h)P(z_0|z_0<z)dz_0=\nonumber \\
&=\dfrac{\int_{-\infty}^{z}L_{av}(z_0, h)w_z^{(1)}(z_0)dz_0}{\int_{-\infty}^{z}w_z^{(1)}(z_0)dz_0}
\end{eqnarray}
where $w_z^{(1)}$ is one-dimensional stationary distribution of process $\mathcal{Z}(x)$ \eqref{Gauss_dens}. 

The case $z>h$ differs from the previous one only by limits of integration
\begin{eqnarray}
&\mathcal{L}^+(z,h)=\int L_{av}(z_0, h)P(z_0|z_0<z)dz_0=\nonumber \\
&=\dfrac{\int_{-\infty}^{h}L_{av}(z_0, h)w_z^{(1)}(z_0)dz_0}{\int_{-\infty}^{z}w_z^{(1)}(z_0)dz_0}
\end{eqnarray}

After substitution of expression for $w_z^{(1)}$ \eqref{Gauss_dens}, one can get
\begin{align}
	\label{L+}
	\mathcal{L}^{-}(z,h) =\sqrt{\dfrac{2}{\pi\sigma^2}}\frac{1}{1+\erf \dfrac{z}{\sqrt{2}\sigma}}\int_{-\infty}^{z}dz_0e^{\frac{-z_0^2}{2\sigma^2}}L_{av}\left(z_0,h\right)
\end{align}
\begin{align}
	\label{L-}
	\mathcal{L}^{+}(z,h) =\sqrt{\dfrac{2}{\pi\sigma^2}}\frac{1}{1+\erf \dfrac{z}{\sqrt{2}\sigma}}\int_{-\infty}^{h}dz_0e^{\frac{-z_0^2}{2\sigma^2}}L_{av}\left(z_0,h\right)
\end{align}
for the cases of $z\leq h$ and $z>h$, respectively:

%$$
%\bar{ L}(z,h) =\frac{1}{\sigma}\frac{\int_{-\infty}^{h}dz_0e^{\frac{-z_0^2}{2\sigma^2}}L\left(z_0,h\right)}{\sqrt{\frac{\pi}{2}}\left(1+\erf \frac{z}{\sqrt{2}\sigma}\right)}
%$$
Analytical integrations for  \eqref{L+}, \eqref{L-}  using exact expression of $L_{av}\left(z_0,h\right)$ \eqref{Average_length_Analytic} is impossible, so approximated expression of $\mathcal{L}\left(z, h\right)$ is needed. Corresponding cumbersome calculations can be found in Appendix~\ref{Integration}. The final approximation for $z<h$ has the following form:
\begin{eqnarray}
\label{L_final_approx-}
&\alpha\mathcal{L}^-_{app}=\alpha l(h)-\left(1+\erf \frac{z}{\sqrt{2}\sigma}\right)^{-1}\times \nonumber\\ 
&\left[\sqrt{\frac{2}{3}}\left(1+\erf\sqrt{\frac{3}{2}}\frac{z}{\sigma}\right)+\left(\frac{1}{\sqrt{2}}A-e^{-\frac{z^2}{2\sigma^2}}\right) \left(1+\erf \frac{z}{\sigma}\right)\right] \nonumber \\
\end{eqnarray}
Auxiliary function $l(t)$ is introduced:
\begin{eqnarray}
&\alpha l(t)=\frac{\pi}{2\sigma}\erfi \frac{t}{\sqrt{2}\sigma}\left(1+\erf \frac{t}{\sqrt{2}\sigma}\right)+ \nonumber \\
&+A\sqrt{\frac{\pi}{2}}\frac{\sigma}{t}e^{-\frac{t^2}{2\sigma^2}}\erfi\frac{t}{\sqrt{2}\sigma}
\end{eqnarray}
where constant $A=2\ln 2/(4-\pi)$. Detailed calculations can be found in Appendix~\ref{Integration}.
%where $\gamma(n,t)$ is incomplete Gamma function. 
%Thus the final approximation for $\mathcal{L}(z,h)$ at the case $z<h$ has the following form:
%\begin{eqnarray}
%\label{L_final_approx+}
%&\alpha\mathcal{L}_{app}=\alpha l(h)
%-\left(1+\erf \frac{z}{\sqrt{2}\sigma}\right)^{-1}\times \nonumber\\ 
%&\left[\sqrt{\frac{2}{3}}\left(1+\erf\sqrt{\frac{3}{2}}\frac{z}{\sigma}\right)+\left(\frac{1}{\sqrt{2}}A-e^{-\frac{z^2}{2\sigma^2}}\right) \left(1+\erf \frac{z}{\sigma}\right)\right] \nonumber \\
%\end{eqnarray}
For the case $z>h$ approximated expression can be written as
\begin{eqnarray}
\label{L_final_approx+}
&\alpha\mathcal{L}^{+}_{app}=\alpha l(h)\dfrac{1+\erf\frac{h}{\sqrt{2}\sigma}}{1+\erf\frac{z}{\sqrt{2}\sigma}}
-\left(1+\erf \frac{z}{\sqrt{2}\sigma}\right)^{-1}\times \nonumber\\ 
&\left[\sqrt{\frac{2}{3}}\left(1+\erf\sqrt{\frac{3}{2}}\frac{h}{\sigma}\right)+\left(\frac{1}{\sqrt{2}}A-e^{-\frac{h^2}{2\sigma^2}}\right) \left(1+\erf \frac{h}{\sigma}\right)\right] \nonumber \\
\end{eqnarray}

\begin{figure}[H]
	\includegraphics[width=8cm]{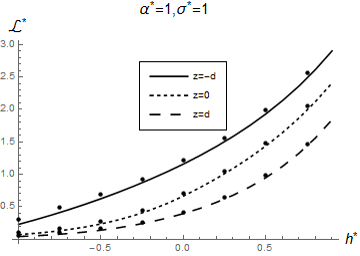}
	\caption{\label{Fig:Final_L} Comparison of numerical results for dimensionless average length $\mathcal{L}^*=\mathcal{L}/d$ and dimensionless  analytical approximations  $\mathcal{L}_{app}^*=\mathcal{L}_{app}/d$.
	Solid and dashed curves correspond to average length $\mathcal{L}_{app}$ calculated from expressions \eqref{L_final_approx+} and \eqref{L_final_approx-} as function of dimensionless $h^*=h/d$ for different positions of the particle $z$ with $\alpha^*=1$, $\sigma^*=1$. Dots correspond to numerical calculations according to \eqref{L+}, \eqref{L-} }
\end{figure}

In the Fig.~\ref{Fig:Final_L} one can find behavior of $\mathcal{L}$ calculated according to approximations \eqref{L_final_approx-}, \eqref{L_final_approx+} in comparison with exact numerical results.

\subsection{Layers Integration}
In this section we perform the calculations of effective fluid-solid potential using obtained results of above sections Let us consider one of the most popular pair intermolecular potential Mie
\begin{eqnarray}
\label{Mie}
U(R)=C \left[ \left(\frac{\sigma}{r}\right)^{\lambda_r}-\left(\frac{\sigma}{r}\right)^{\lambda_a}\right],
\end{eqnarray}
where $R$ is the distance between molecules, $C=\frac{ \lambda_r}{\lambda_r-\lambda_a}\left(\frac{\lambda_r}{\lambda_a}\right)^{\frac{\lambda_r}{\lambda_r-\lambda_a}}$ is a constant. In case of Lennard-Jones (LJ) fluid ($\lambda_r=12, \lambda_a=6$) this constant equals to $C=4\epsilon$, where $\epsilon$ is characteristic energy. One can consider a term of above expression as general power function of $R$ in the following form $C d^{\gamma}/R^{\gamma}$. The interaction energy of molecule and surface, induced by pair potential $U(R)$ is the sum of interactions with all molecules in the solid media. 

It is well known, that in the case of planar surface the integration can be easily obtained. In cylindric system of coordinate, circular ring of solid media  with radius $r$ and width $dz$ has the volume $2\pi rdz dr$. The number of solid molecules in the ring will be $2\pi\rho_s rdzdr$, where $\rho_s$ is the number density of solid molecules. Then the interaction energy for a molecule at a distance $D$ away from the surface is integration over all solid media
\begin{eqnarray}
\label{Part_potential_flat}
&U^{(\gamma)}(D)=2\pi C\rho_s\int\limits_{D}^{\infty}dz\int\limits_{0}^{\infty}\dfrac{rdr}{\left(z^2+r^2\right)^{\gamma/2}}=\nonumber \\
&=\dfrac{2\pi C\rho_s}{(\gamma-2)(\gamma-3)D^{\gamma-3}} \nonumber
\end{eqnarray}
In the case of Lennard Jones potential, one can get well known expression:
\begin{equation}
\label{9_3_potential}
U_{9-3}(D)=2\pi\rho_s\epsilon \sigma^3\left[\frac{2}{45}\left(\frac{d}{D}\right)^{9}-\frac{1}{3}\left(\frac{d}{D}\right)^{3}\right]
\end{equation}
Obviously, in the case of geometry heterogeneous surface above calculations becomes inadequate and modifications are needed. Firstly, in this case spatial density of solid is not constant and is defined by $\rho_s P_{11}(r,h)$, see result of \ref{Probability}. Secondly, spatial $r$-integration at each layer $h$ should be started from corresponding value of $\mathcal{L}(h,z)$ see result of \ref{Average}. Thus, the new formula for the number of molecules in the ring of solid surface is $2\pi\rho_s P_{11}(z,r-\mathcal{L}) rdzdr$, where integration over $r$ will be started from  $\mathcal{L}(h,z)$. Thus, integration of $U^{(\gamma)}$ over certain layer with $z<h$ is defined as:
 
 \begin{equation}
\label{Heter_Pot_Layer_1}
U_{layer}^{(\gamma)}(\mathcal{L};h,z)=2\pi C\rho_s\int_{\mathcal{L}}^{\infty}rdr\frac{P_{11}(r-\mathcal{L},h)}{\left[\left(z-h\right)^2+r^2\right]^{\gamma/2}}
\end{equation}
After the following substitution $r'=r-\mathcal{L}$ the above integration starts from zero
\begin{equation}
\label{Heter_Pot_Layer_2}
U_{layer}^{(\gamma)}(\mathcal{L};h,z)=2\pi C\rho_s\int_{0}^{\infty}r'dr'\frac{P_{11}(r',h)}{\left[\left(z-h\right)^2+\left(r'+\mathcal{L}\right)^2\right]^{\gamma/2}}
\end{equation}
Result corresponding to Mie potential \eqref{Mie} can be obtained as
\begin{eqnarray}
\label{U_layer}
U_{layer}(\mathcal{L};h,z)=U_{layer}^{(\lambda_r)}(\mathcal{L};h,z)-U_{layer}^{(\lambda_a)}(\mathcal{L};h,z)
\end{eqnarray}
%For any values of $r, h$ the function $P_{11}(r,h)$ is confined and has the largest value near $r=0$ and then with increasing of $r$ very quickly tends to constant value. Thus, from the structure of above integral it is sufficient to expand $P_{11}(r,h)$ near $r=0$:
%\begin{equation}
%\label{P11_Approx}
%P_{11}(x,h)=%C_0+\frac{1}{C_0}e^{\tilde{h}^2}T\left(\sqrt{2}\tilde{h};1\right)
%1-\frac{1}{2\pi C_0}\sqrt{\frac{\alpha r}{2}}
%\end{equation}
%After the integration with \eqref{P11_Approx} one has:
%\begin{equation}
%\begin{split}
%\label{Heter_Pot_Layer_Approx}
%2\pi C\rho_s \left[\frac{\left[\zeta^2+L^2\right]^{1-\frac{\gamma}{2}}}{\gamma-2}-\frac{L^{2-\gamma}}{\gamma-1} {}_2F_1\left(\frac{-1+\gamma}{2},\frac{\gamma}{2},\frac{1+\gamma}{2},-\frac{\zeta^2}{L^2}\right)-\right.  \\
%\left. -\frac{1}{2\pi C_0}\sqrt{\frac{\alpha}{2}}\frac{3L^{\frac{5}{2}-\gamma}\sqrt{\pi}}{4\Gamma(\gamma)}{}_2F_1\left(\frac{-5+2\gamma}{4},\frac{-3+2\gamma}{4},\frac{1+\gamma}{2},-\frac{\zeta^2}{L^2}\right)\right]
%\end{split}
%\end{equation}
The total interaction potential of a molecule and the solid media is the integration of \eqref{U_layer} over $h$ from $-\infty$ to $\infty$. Thus, desired general expression has the following form:
\begin{eqnarray}
\label{General_Result}
U^{eff}_{fs}(z)=\int\limits_{-\infty}^{z}dhU_{layer}(\mathcal{L}^{+};h,z)+\int\limits_{z}^{\infty}dhU_{layer}(\mathcal{L}^{-};h,z) \nonumber \\
\end{eqnarray}
Thus, obtained result contains two explicit integrals and can be calculated using results for $P_{11}$ \eqref{P11_result}, $\mathcal{L}^-_{app}$ \eqref{L_final_approx-} and $\mathcal{L}^+_{app}$ \eqref{L_final_approx+}.

\section{Results and Discussion}

In this section obtained expression \eqref{General_Result} is applied to various heterogeneous surfaces which can be described by random process with parameters $\tau, \sigma$. As an example we considered two groups of surfaces: the first one is amorphous materials with structureless surface and high geometric heterogeneity; the second one is materials with well-defined molecular structure and small heterogeneity.   

\begin{figure}[H]
	\begin{subfigure}{0.5\textwidth}
		\includegraphics[width=8.5cm]{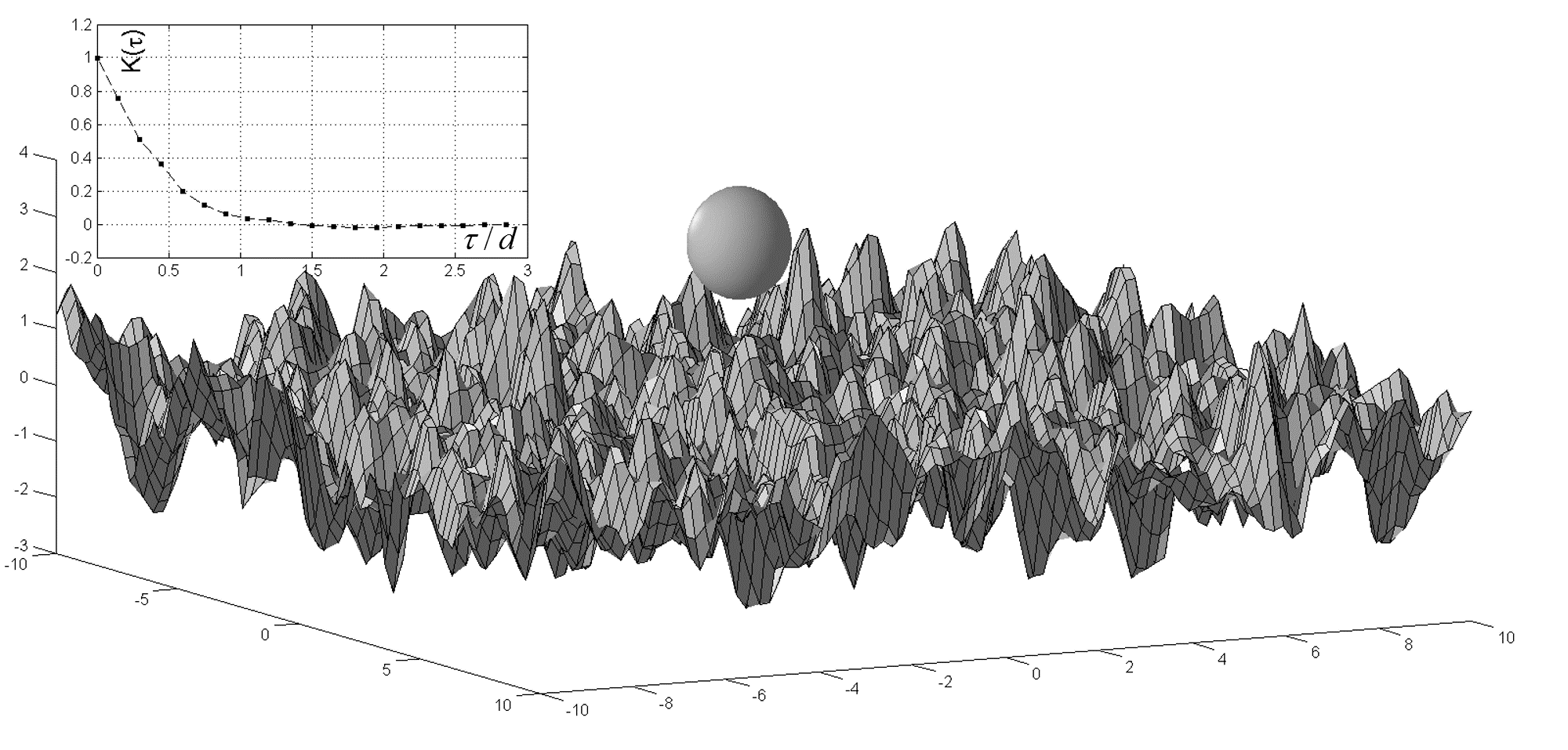}
		\caption{The surface with $\sigma=0.8 d$, $\tau=0.5 d$}
		\label{fig:HeterLargeExamples1}
	\end{subfigure}
	\qquad 
	\begin{subfigure}{0.5\textwidth}
		\includegraphics[width=8.5cm]{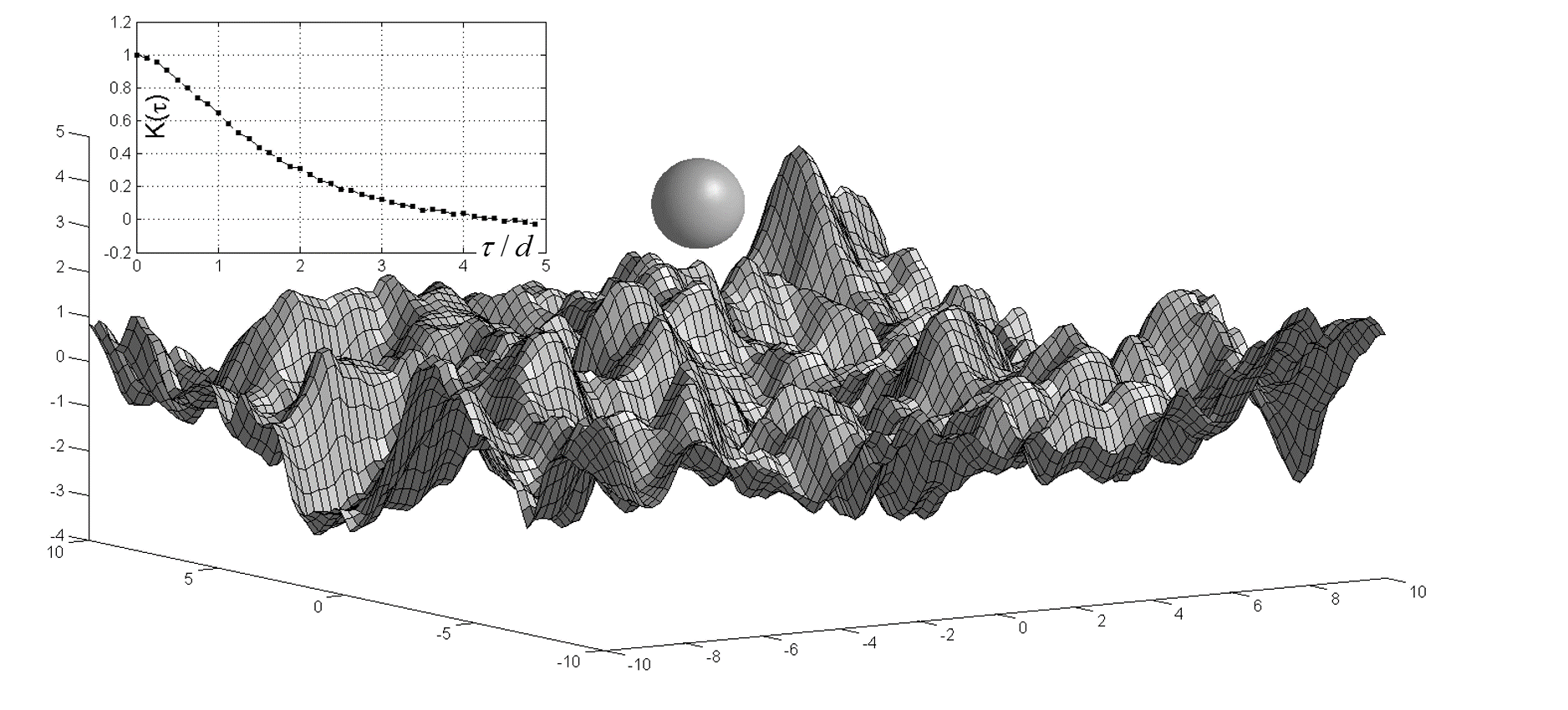}
		\caption{The surface with $\sigma=1 d$, $\tau=1.5 d$}
		\label{fig:HeterLargeExamples2}
	\end{subfigure}
	\caption{Examples of simulated heterogeneous surfaces with fixed parameters $\sigma$, $\tau$. In the insets correlation functions $K(x)$ are presented.}
	\label{fig:HeterLargeExamples}
\end{figure}

\subsection{Amorphous materials with high heterogeneity}

The examples of  heterogeneous surfaces generated by random process simulations with fixed $\tau, \sigma$ can be found in Fig.~\ref{fig:HeterLargeExamples}. 
In this terms ideal plane surface corresponds to random process with $\sigma=0$. Let us define this ideal limit as zero-plane. In spite of similar variances $\sigma$ effective molecular interaction can be quite different due to significant differences of correlation lengths $\tau$. Indeed from Fig.~\ref{fig:HeterLargeExamples1} one can see, that surface has very sharp structure, since, peaks of the surface are located close together. This structure prohibits a particle to approach the zero-plane. Another case with smoother surface structure where fluid particle can easily pass throw the zero-plane  is shown in Fig.~\ref{fig:HeterLargeExamples2}.
These qualitative discussions can be illustrated by calculation of effective potential according to expression \eqref{General_Result}.    
\begin{figure}[H]
	\includegraphics[width=8cm]{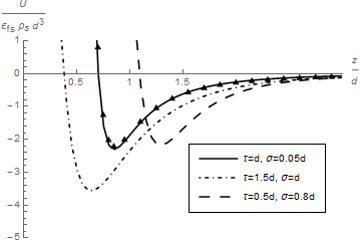}
	\caption{Solid, dash and dash-dotted lines correspond to dimensionless energy potential as function of dimensionless distance $z/d$ for three surfaces, obtained from \eqref{General_Result}. Triangles correspond to classical 9-3 Lennard-Jones potential for ideal smooth surface \eqref{9_3_potential}.}
	\label{fig:HeterPotGeneral}
\end{figure}

In Figure \ref{fig:HeterPotGeneral} solid line is potential of flat, smooth surface, and this curve coincides with known expression \eqref{9_3_potential} (triangles in the figure). Dashed and dash-dotted lines refer to energy potentials for surfaces from Fig.~\ref{fig:HeterLargeExamples}, these curves are quite different and their behavior stay in agreement with above discussions. Dashed line corresponds to the surface from Fig.~\ref{fig:HeterLargeExamples1}, the position of minimum is shifted to the right from the case of flat one due to sharp heterogeneity. Dash-dotted line is the case from Fig.~\ref{fig:HeterLargeExamples2} where the minimum of effective interaction potential is shifted to the left. Thus the molecules in the case from Figure \ref{fig:HeterLargeExamples2} (dash-dotted line) can pass deeper in the solid media than in the case of flat surface. For the surface \ref{fig:HeterLargeExamples1} (dash line) the opposite situation takes place, as it was discussed above. 

\subsection{Materials with small heterogeneity}

\subsubsection{Small scale approximation}
In the case of materials with small variance of the surface, general formula for effective fluid-solid interaction potential may be significantly simplified. One can use the observation that near $z=0$ the second part of the integral \eqref{General_Result} is much smaller than the first one.

\begin{figure}[H]
	\includegraphics[width=9cm]{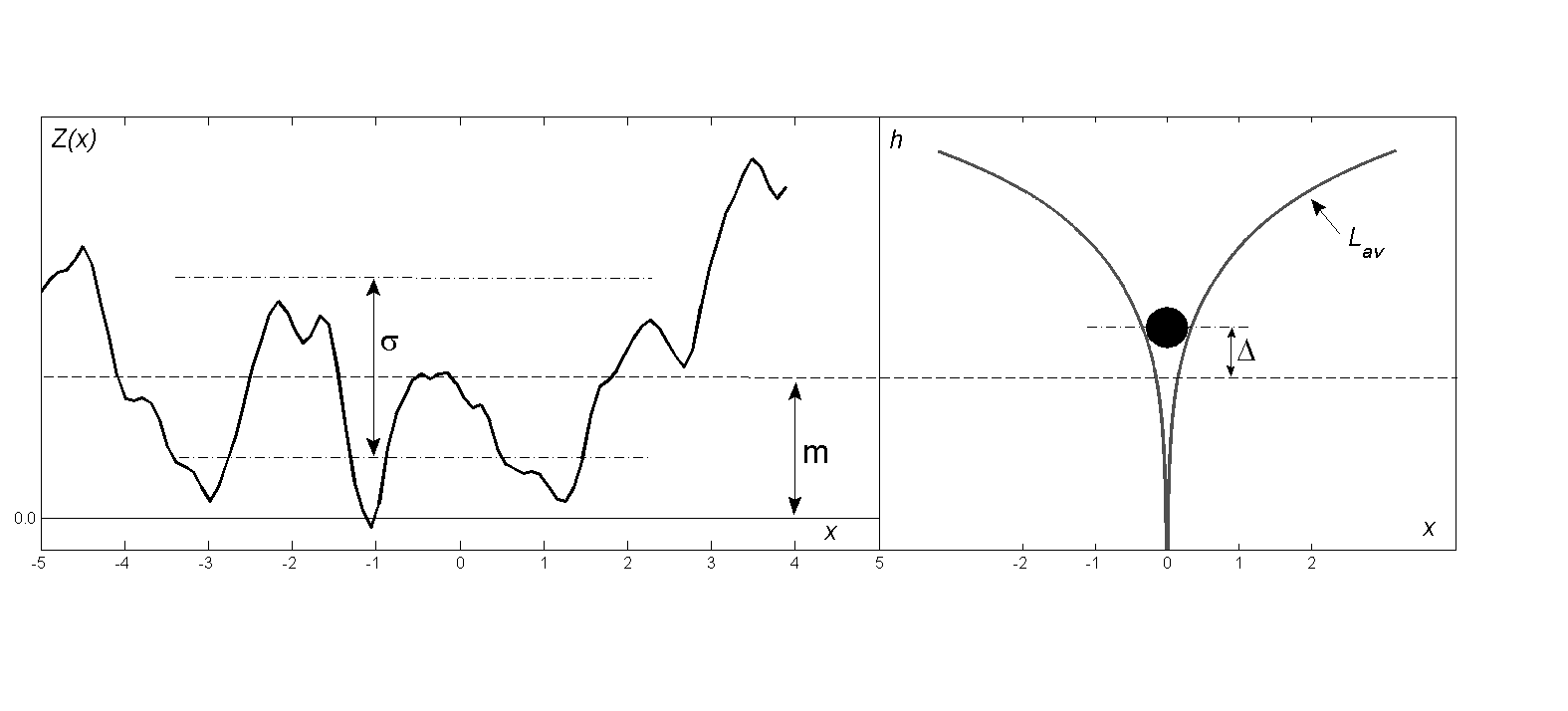}
	\caption{Left figure illustrates some realization of random process $Z(x)$ which describes the solid surface. The mean value $m$ and variance $\sigma$ are presented. On the right side of the figure $\mathcal{L}$ function and corresponding minimal distance $\Delta$ are demonstrated.}
	\label{fig:Delta}
\end{figure}

Let us consider the $\mathcal{L}(z,h)$ function, which in essence is the average behavior of the random solid in each layer $h$ in presence of fluid particle at the distance $z$ from the mean value $m$ of the solid surface. For the surfaces with small $\sigma$ the value of $\mathcal{L}(z,h)$ increases rapidly on the interval $z>0$. Moreover, in the limit $\sigma\to 0$ this function becomes zero for $z<0$ and infinity for $z>0$. Thus, the point at the first contact of the fluid particle with diameter $d$ and the curve  $\mathcal{L}(z,h)$ corresponds to the minimal distance $\Delta$ between center of particle and mean value $m$. Graphical explanation can be found in Fig.\ref{fig:Delta}. Existence of the minimal distance $\Delta$ is caused by repulsion hard sphere potential between fluid and solid particles.
Thus in this case the general formula for effective potential may be simplified and has the following form
\begin{eqnarray}
\label{Ufs_App_1}
U_{fs}^{App}(z)=\int_{-\infty}^{\Delta}U_{layer}(L^{+}(z,h),z,h)dh. 
\end{eqnarray}

Furthermore one more step of simplification could be done. One can replace the value of integral \eqref{Ufs_App_1} by interaction potential of some reference system with appropriate shifting. In this work the reference system corresponds to solid media with flat surface and is described by well Steelle potential \cite{steele1973physical}. 
\begin{eqnarray}
&U_{St}(z)=2\pi \rho_s\epsilon_{sf}\sigma_{sf}^2\delta\left[\left(\dfrac{\sigma_{sf}}{z}\right)^{10}-\left(\dfrac{\sigma_{sf}}{z}\right)^{4}+ \nonumber \right. 
\\
&\left. +\dfrac{\sigma_{sf}^4}{3\delta(0.61\delta+z)^3}\right] 
\end{eqnarray} 

where $\rho_s=0.114 \AA^{-3}$ is the number density of carbon atoms in graphite, $\delta=3.35\AA$ is the interlayer spacing in graphite, $\epsilon_{sf}$ and $\sigma_{sf}$ are the characteristic energetic and scale parameters of the solid-fluid LJ potential.
According to general property of interaction potential the point where $U_{fs}^{App}(z)$ becomes zero corresponds to the minimal distance $\Delta$. In other hand in the case of Steele potential this point $z^{\star}$ can be obtained easily from explicit equation $U_{St}(z^{\star})=0$. Thus approximation of effective fluid-solid potential can be written as

\begin{eqnarray}
&U_{fs}^{App}(z)=U_{St}(z-\Delta+z^{\star})
\end{eqnarray}

\begin{widetext}
	\begin{figure}[H]
		\includegraphics[width=18cm]{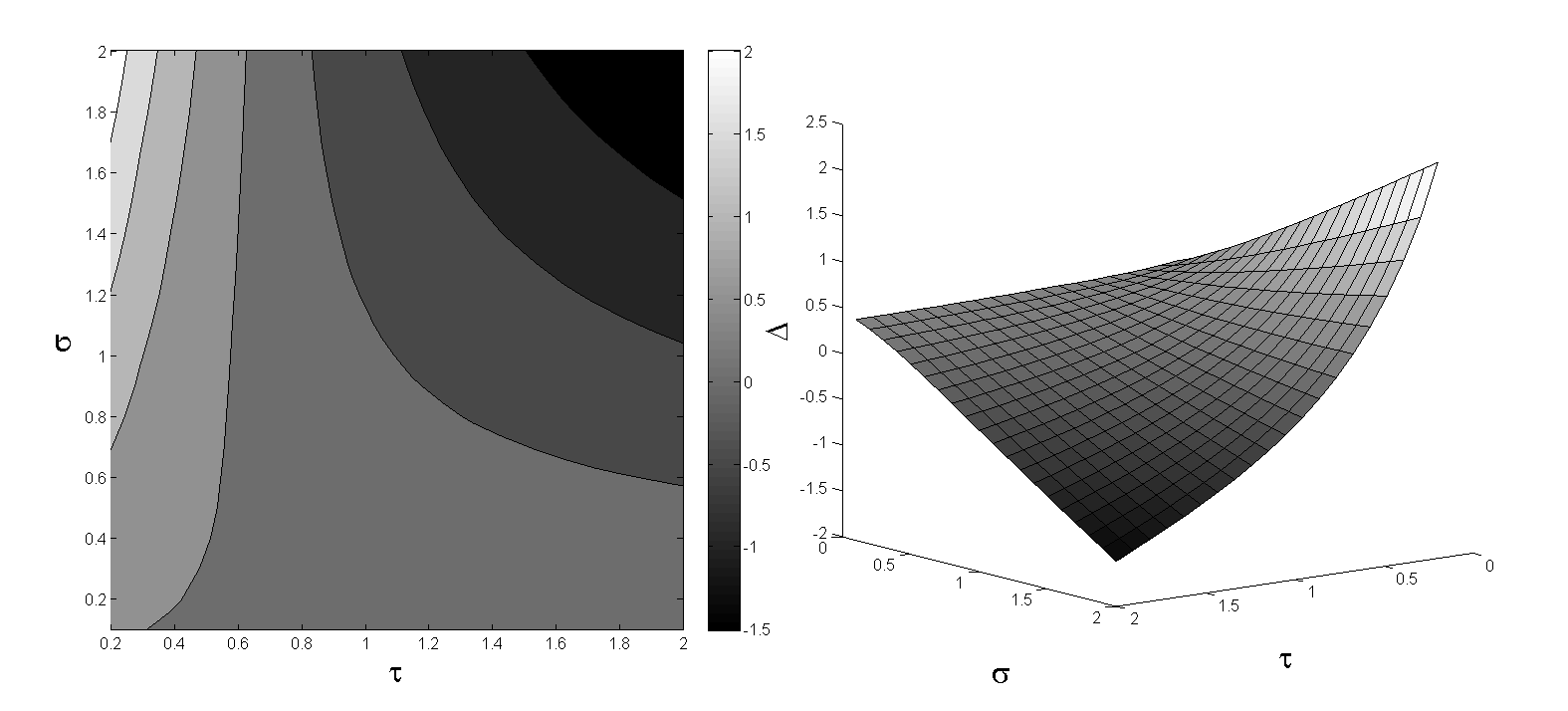}
		\caption{Penetration length $\Delta$ as a function of relative values of variance $\sigma/d$ and correlation length $\tau/d$ of the surface obtained by numerical calculations.}
		\label{fig:Delta_Isolines}
	\end{figure}
\end{widetext}

It is important to note that $\Delta$ is strongly depends on the variance $\sigma$, correlation length $\tau$ of the surface and fluid molecular diameter $d$. Results of numerical calculations which demonstrate behavior of $\Delta$ can be found in Fig.~\ref{fig:Delta_Isolines}

It follows From Fig.~\ref{fig:Delta_Isolines}, that in the case of very small geometric heterogeneity ($\sigma$ near zero) lateral structure of the surface is not important. However with increasing the value of $\sigma$ influence of correlation length $\tau$ becomes rapidly crucial. Thus one can see that small (smaller than molecule diameter) values of $\tau$ lead to effective repulsion of fluid particle from the solid due to geometric effects. In contrast large values of $\tau$ cause the deeper penetration of fluid particle in solid media.

\subsubsection{Well-defined materials}

\begin{widetext}
	\begin{figure}[H]
		\includegraphics[width=17cm]{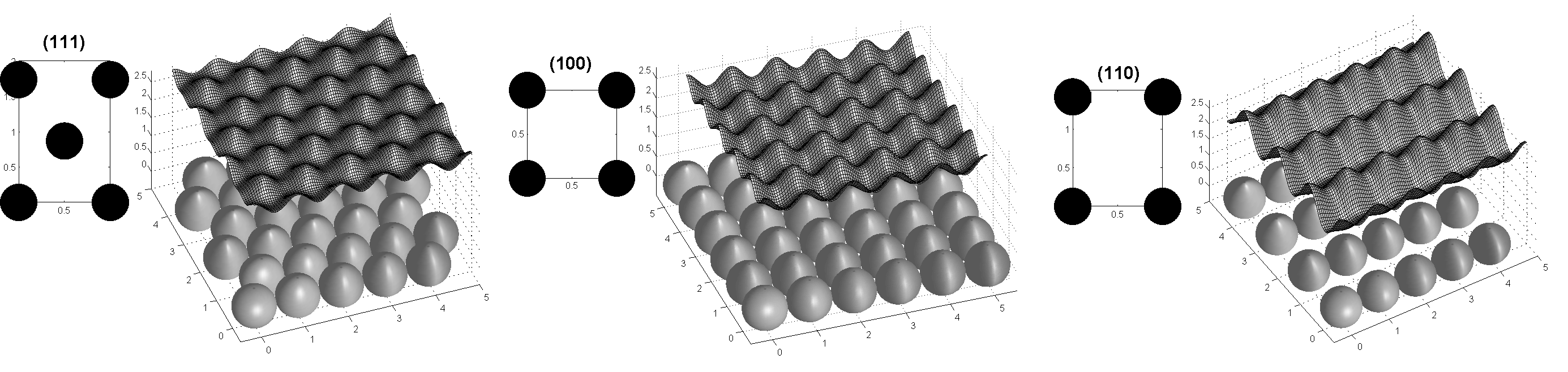}
		\caption{Illustration of three types of fcc lattice (111), (100), (110). Each picture consists of lattice face scheme, molecular structure pattern and corresponding covered surface.}
		\label{fig:Patterns}
	\end{figure}
\end{widetext}

Results of this work can be formally used to describe of molecular interaction with well defined materials. Let us consider one of the main type of cubic crystals -- face centered cubic (fcc) ones. More precisely crystals with the following surface lattices (111), (100), (110) are considered. Schematic illustration of these cases can be found in Fig~\ref{fig:Patterns}. 

\begin{figure}[H]
		\centering
		%\captionsetup{width=2\linewidth			}
		\includegraphics[width=8.5cm]{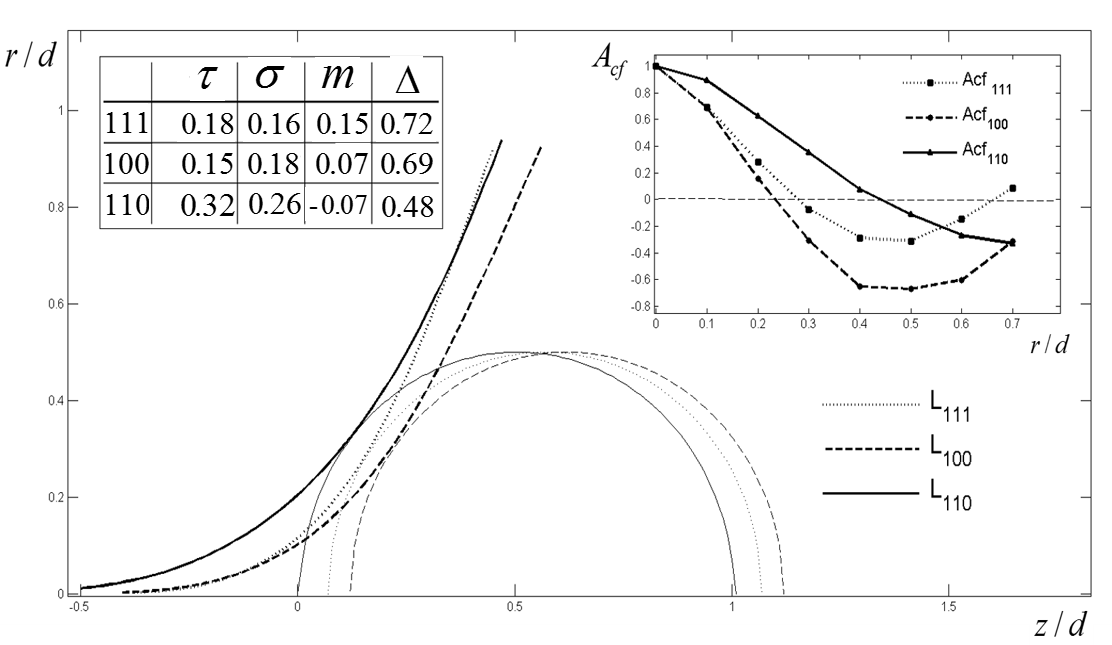}
		\caption{Major figure illustrates condition on the first contact of spherical molecule and averaged profile of heterogeneous surface for three types of fcc (111), (100), (110). Table shows characteristic parameters for considering surfaces, inset shows corresponding autocorrelation functions obtained numerical.}
		\label{fig:StructureSurfResult1}
	\end{figure}

Obviously, the boundary surface of these materials is not smooth on the molecular scale. As one can see from Fig~\ref{fig:Patterns} each molecular structure can be approximated by heterogeneous surface. It is important to note, that these surfaces are deterministic and direct analogy with above results is not correct. However instead of averaging over random process realizations, it is possible to use spatial averaging over the surface geometry. Then parameters $\sigma, \tau, m, \Delta$ which characterize geometry of the surface and can be calculated numerically. Numerical results can be found in Fig.~\ref{fig:StructureSurfResult1}. As one can see $\Delta$ strongly depends from surface type.
%\begin{widetext}

\begin{figure}[H]
  	\includegraphics[width=8.5cm]{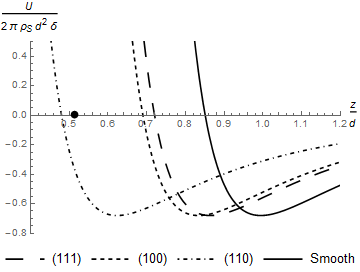}
  	\caption{Effective fluid-solid potentials corresponding to small scale approximation. The black dot corresponds to value of $\Delta$ from Monte-Carlo simulations \cite{forte2014effective} for (110) solid media.}
  	\label{fig:StructureSurfResult2}
  \end{figure}

Obtained effective fluid-solid potentials corresponding to small scale approximation can be found in Fig.~\ref{fig:StructureSurfResult2}. One can see that all three potentials are shifted to the left in comparison with smooth case (Steele potential corresponds to solid line in the figure). This fact illustrates that heterogeneous potential in these cases is softer than smooth one.  Considered cases have the following order: (110) has the softest potential (the smallest $\Delta$), then (100) and (111), respectively. This order agrees well with results of Monte-Carlo simulations \cite{forte2014effective} for monolayers with the same fcc structures. Obtained results correlate with atomic patterns schemes Fig~\ref{fig:Patterns}. Indeed case (111) has the most dense packing stricture which prohibits molecular penetration, otherwise in case (110) structure is not so close and fluid molecules can penetrate deeper. 

%Whole curve of (110) potential from \cite{forte2014effective} not shown here due to tempreta
%Numerical modeling of averaged potential for considered here fcc can be found in work \cite{forte2014effective}. This published result has temperature dependence and for this reason direct numerical comparison with obtained in our work results is not correct. However, the minimal position of molecule center, in other words zero of potential $u(\Delta)=0$ can be compared. Curves in Fig~\ref{fig:Patterns} are calculated relative to value of minimum for corresponding smooth potential.
\section{Conclusion}

In current research we developed the general theory of effective fluid-solid potential for heterogeneous surfaces. Our approach is based on theory of Markovian random processes and the first passage time probability problem. The first passage time probability problem was reformulated and applied as part of averaging procedure. It can be applied for wide range  of correlation functions of the random solid surface. 

Finally numerical results for effective fluid-solid potentials in the case of amorphous materials with high heterogeneity were obtained. Also the general formula for potentials was simplified in the case of small surface heterogeneity. Obtained expression was applied to several types of face cubic centered crystals. It was shown that the wider is the lattice spacing in terms of molecular diameter of the fluid, the more different is between obtained potentials and homogeneous one. Also this effect was demonstrated in \cite{forte2014effective} by fully atomistic Monte-Carlo simulations. A comparison is presented that shows good qualitative agreement of theory predictions and simulation. The method provides a promising approach to explore how the random geometry heterogeneity effect on thermodynamic properties of the fluid which is highly desirable in any DFT calculations.
\appendix
\section{\label{A1} Condition probability}
%\subsection{Condition probability}
The first step is calculation of the sum in \eqref{Probability11}. Let us start with the follow expression for Hermits polynomials:
\begin{align}
	\label{ChrisDarbu}
	\sum_{n=0}^{\infty}\frac{1}{n!}H_n^2\left(\tilde{h}\right)\left(\frac{u}{2}\right)^n
	=\frac{1}{\sqrt{1-u^2}}\exp\left(\frac{2u}{1+u}\tilde{h}^2\right)
	\end{align}
where $\tilde{h}=h/\sigma$. After simple modification in \eqref{Probability11} one can rewrite initial sum using \eqref{ChrisDarbu}:
	\begin{align}
		\label{P11Sum1}
		\sum_{n=1}^{\infty}\frac{1}{n!}H_{n-1}^2\left(\tilde{h}\right)\left(\frac{K}{2}\right)^n= \nonumber \\
		=-\frac{1}{2}e^{\tilde{h}^2}\int_{\pi/2}^{\phi}d\phi'\exp\left(-\tan^{2}\left(\frac{\phi'}{2}\right)\tilde h^2\right)
		=-\frac{1}{2}e^{\tilde{h}^2}I
		\end{align}
where $\phi=\arccos K$. Integrand of $I$ can be expanded in Taylor series around zero: 
		\begin{align}
			\label{I_Taylor}
			I=\phi-\frac{\pi}{2}
			+\int_{\pi/2}^{\phi}d\phi'\sum_{n=1}^{\infty}\frac{(-1)^{n}\tilde{h}^{2n}}{n!}\left(\tan\frac{\phi'}{2}\right)^{2n}
			\end{align}
For integration the following formula is more convenient:
			$$
			\int \tan^{2n} x dx= (-1)^n x+\sum_{k=1}^{n}(-1)^{k-1}\frac{\tan^{2n-2k+1}x}{2n-2k+1} 
			$$
after integration in \eqref{I_Taylor} one can obtain
			\begin{eqnarray}
			&I=\left(e^{\tilde{h}^2}-1\right)\arcsin K - \nonumber \\
			&-2\sum\limits_{n=1}^{\infty}\frac{(-1)^{n}\tilde{h}^{2n}}{n!}\sum\limits_{k=1}^{n}\frac{(-1)^{k-1}}{2n-2k+1}\left[\left(\tan\frac{\phi}{2}\right)^{2(n-k)+1}-1\right] \nonumber
			\end{eqnarray}
in the last sum one can make summation over new index $j=n-k+1$
			\begin{eqnarray}
			\sum_{k=1}^{n}\frac{(-1)^{k-1}}{2n-2k+1}\left[\left(\tan\frac{\phi}{2}\right)^{2(n-k)+1}-1\right]=
			\nonumber \\
			=\sum_{j=1}^{n}\frac{(-1)^{n-j}}{2j-1}\left[\left(\tan\frac{\phi}{2}\right)^{2j-1}-1\right] \nonumber
			\end{eqnarray}
			after substitution $y=\tan \frac{\phi}{2}$ above result can be written as
			\begin{eqnarray}
			\frac{1}{2}(-1)^{n}\arcsin K+\frac{y^{2n+1}}{2n+1}{}_2F_1\left(1;\frac{1}{2}+n;\frac{3}{2}+n;-y^2 \right)-\nonumber\\
			-\frac{1}{2n+1}{}_2F_1\left(1;\frac{1}{2}+n;\frac{3}{2}+n;-1 \right) \nonumber
			\end{eqnarray}
			where ${}_2F_1$ is hypergeometric function \cite{bateman1955higher}. There is integral representation for  ${}_2F_1$ in general case:
			\begin{align}
				\label{def_hypergoem_Lerch}
				{}_2F_1\left(a;b;c;z\right)=\frac{\Gamma\left(c\right)}{\Gamma\left(b\right)\Gamma\left(c-b\right)}\int_{0}^{1}\frac{t^{b-1}(1-t)^{c-b-1}}{\left(1-tz\right)^a} 
				\end{align}
				%\Phi\left(z,s,a\right)=\sum_{k=0}^{\infty}\frac{z^k}{\left(a+k\right)^s}
				In order to calculate $I$ one can use expression \eqref{def_hypergoem_Lerch}, where summation over $n$ becomes:  
				
				\begin{eqnarray}
				I=-\arcsin K- \nonumber \\
				-\sum_{n=1}^{\infty}\frac{(-1)^{n}\tilde{h}^{2n}}{n!}\left[y^{2n+1}\int_{0}^{1}\frac{ t^{n-\frac{1}{2}}dt}{1+y^2 t}-\int_{0}^{1}\frac{ t^{n-\frac{1}{2}}dt}{1+ t}\right]=\nonumber \\
				=-\arcsin K+J\left(\tilde{h},y\right)-J\left(\tilde{h},1\right) \nonumber
				\end{eqnarray}
				where the following expression has been used:
				\begin{eqnarray}
				J\left(\tilde{h},y\right)=-\int_{0}^{1}\frac{ y t^{-\frac{1}{2}}dt}{1+y^2 t}\sum_{n=1}^{\infty}\frac{(-1)^{n}\left(\sqrt{t}y\tilde{h}\right)^{2n}}{n!}=\nonumber \\
				=-\int_{0}^{1}\frac{ y t^{-\frac{1}{2}}dt}{1+y^2 t}\left(e^{-\left(\sqrt{t}y\tilde{h}\right)^2}-1\right)= \nonumber \\
				=2\int_{0}^{y}\frac{ dq}{1+q^2 }e^{-\left(q\tilde{h}\right)^2}-2\arctan y=\nonumber \\
				=4\pi e^{\tilde{h}^2}T\left(\sqrt{2}\tilde{h},y\right)-2\arctan y \nonumber
				\end{eqnarray}
			where Owen's T function is defined as \cite{owen1956tables}
				\begin{align}
					\label{OwenT_1}
					T\left(\tilde{h},a\right)=\frac{1}{2\pi}\int_{0}^{a}\frac{e^{-\frac{1}{2}\tilde{h}^2\left(1+x^2\right)}}{1+x^2}dx
					\end{align}
					Now one can write
					\begin{align}
						\label{I_result}
						I=4\pi e^{\tilde{h}^2}\left[T\left(\sqrt{2}\tilde{h};y\right)-T\left(\sqrt{2}\tilde{h};1\right)\right]
						\end{align}
						
						In the result there is exact expression for condition probability function \eqref{Probability11}
						\begin{align}
							\label{Probability11_result}
							P_{11}\left(x,h\right)=C_0-\frac{2}{C_0}\left[T\left(\sqrt{2}h/\sigma, y(x)\right)-T\left(\sqrt{2}h/\sigma, 1\right)\right]
							\end{align}

\section{\label{Simplification} Simplification of average length}
 In Appendix \ref{Simplification} some simplifications of the  average length $L_{av}(z_0,h)$ are considered. The average length has the following expression \eqref{Average_length}:
 \begin{align}
 	\label{B1}
 	L_{av}\left(z_0,h\right)=\frac{1}{\alpha \sigma^2}\int_{z_0}^{h}e^{\frac{\xi^2}{2\sigma^2}}d\xi\int_{-\infty}^{\xi}d\eta e^{-\frac{\eta^2}{2\sigma^2}}
 \end{align}
 
 Firstly, the integration in \eqref{B1} over $\eta$ can be splitted into two parts: 
 \begin{align}
 	\label{Average_length_1}
 	&\alpha L_{av}\left(z_0,h\right)=
 	\frac{1}{\alpha \sigma^2}\int_{z_0}^{h}e^{\frac{\xi^2}{2\sigma^2}}d\xi\int_{0}^{\xi}d\eta e^{-\frac{\eta^2}{2\sigma^2}} \\ \nonumber
 	&+\frac{\pi}{2}\left[\erfi\frac{h}{2 \sigma}-\erfi\frac{z_0}{2 \sigma}\right]
 \end{align} 
 
 Secondly, the integration over $\xi$ can be considered in complex plane $\xi\to i y$: 
 \begin{align}
 	\label{Average_length_2}
 	\int_{z_0}^{h}e^{\frac{\xi^2}{2\sigma^2}}d\xi\int_{0}^{\xi}d\eta e^{-\frac{\eta^2}{2\sigma^2}}=i \int_{i z_0}^{i h}e^{-\frac{y^2}{2\sigma^2}}dy\int_{0}^{i y}d\eta e^{-\frac{\eta^2}{2\sigma^2}}
 \end{align}
 In this form there is connection with the expression for Owen's T functions:
 \begin{align}
 	\label{OwenT_2}
 	T(h,a)-T(x,a)=-\frac{1}{2\pi}\int_{x}^{h}\int_{0}^{a y}e^{-\frac{y^2+\eta^2}{2}}dyd\eta
 \end{align}
 One can rewrite integrations \eqref{Average_length_2} with analytical continuation of \eqref{OwenT_2} to complex plane:
 \begin{align}
 	\label{Average_length_3}
 	i \int_{i z_0}^{i h}e^{-\frac{y^2}{2\sigma^2}}dy\int_{0}^{i y}d\eta e^{-\frac{\eta^2}{2\sigma^2}}
 	=-2\pi i\left[T(i h/\sigma,i)-T(i z_0/\sigma,i)\right]
 \end{align}
 Analytical continuation of Owen's function can be obtained from the definition \eqref{OwenT_1}:
 
 Thus, desired expression for average length has the following form:
\begin{eqnarray}
 	\label{Average_length_5}
 	&\alpha L_{av}\left(z_0,h\right)=\int_{0}^{1}d\tau\dfrac{e^{\frac{1}{2\sigma^2}h^2(1-\tau^2)}-e^{\frac{1}{2 \sigma^2}z_0^2(1-\tau^2)}}{1-\tau^2}+ \nonumber \\
 	&+\frac{\pi}{2}\left[\erfi\dfrac{h}{\sqrt{2}}-\erfi\dfrac{z_0}{\sqrt{2}\sigma}\right]
\end{eqnarray}

\section{\label{Integration} Integration of average length}

The approximation of $\mathcal{L}(z, h)$ in the case of $z>h$ is discussed, the opposite case $z<h$ can be described in the similar way. There are two main characteristics of approximation $\mathcal{L}_{app}(z, h)$. Firstly, for accurate numerical calculations, approximated expression a point (0, 0) has to be equal to exact one $\mathcal{L}_{app}(0, 0)=\mathcal{L}(0, 0)$.  Secondly, for right physical meaning, exact integration at $z0\to-\infty$ is needed. For this reason correct limit of $L_{av}\left(z_0,h\right)$ at $z0\to-\infty$ is used. 

According to definition of $L_{av}\left(z_0,h\right)$ \eqref{Average_length} consider the following indefinite integral:
\begin{eqnarray}
&\sqrt{\frac{\pi}{2}}\int\frac{1}{\sigma} dte^{\frac{t^2}{2}}(1+\erf\frac{t}{\sqrt{2}})=\nonumber \\
&=\frac{\pi}{2}\erfi \frac{t}{\sqrt{2}\sigma}\left(1+\erf \frac{t}{\sqrt{2}\sigma}\right) +\sum_{n=0}^{\infty}\frac{\Gamma\left(n+1,\frac{t^2}{2\sigma^2}\right)}{(2n+1)n!} \nonumber
\end{eqnarray}

Let us start from asymptotic in $t\to-\infty$, for the last term we have $\Gamma(n+1,\frac{t^2}{2\sigma^2})\sim\left(\frac{t^2}{2\sigma^2}\right)^ne^{-\frac{t^2}{2\sigma^2}}$, then
$$
\sum_{n=0}^{\infty}\frac{\Gamma\left(n+1,\frac{t^2}{2\sigma^2}\right)}{(2n+1)n!}\sim \sqrt{\frac{\pi}{2}}\frac{\sigma}{t}e^{-\frac{t^2}{2\sigma^2}}\erfi\frac{t}{\sqrt{2}\sigma}
$$
Thus $L_{av}$ can be written as:
\begin{eqnarray}
\label{C1}
&L_{av}(z_0,h)\simeq l(h)-l(z_0) \nonumber \\
&\alpha l(t)=\dfrac{\pi}{2\sigma}\erfi \dfrac{t}{\sqrt{2}\sigma}\left(1+\erf \dfrac{t}{\sqrt{2}\sigma}\right)+ \\
&+A\sqrt{\dfrac{\pi}{2}}\dfrac{\sigma}{t}e^{-\frac{t^2}{2\sigma^2}}\erfi\dfrac{t}{\sqrt{2}\sigma} \nonumber
\end{eqnarray}
where constant $A$ is defined from condition $\mathcal{L}_{app}(0, 0)=\mathcal{L}(0, 0)$. The value of $\mathcal{L}(0, 0)$ can be found exactly from definition: 
\begin{eqnarray}
\label{L00_def}
 &\mathcal{L}(0,0)=\sqrt{\frac{2}{\pi}}\int_{-\infty}^{0}dz_0 \times \nonumber \\
&\times \left[\int_{0}^{1}dt\frac{e^{-z_0^2/2}-e^{-z_0^2t^2/2}}{1-t^2}- 
\frac{\pi}{2}e^{-\frac{z_0^2}{2}}\erfi\frac{z_0}{\sqrt{2}}\right]
\end{eqnarray}
To avoid the singularities in calculations principal value of integral is used.
\begin{equation}
p.v.\int_{0}^{\infty}\frac{e^{-z_0^2t^2/2}}{1-t^2}=\frac{\pi}{2}e^{z_0^2/2}\erfi\frac{|z_0|}{\sqrt{2}} 
\end{equation}
%Important to note that integration  \eqref{L00_def} carry out in negative range of $z_0$, for this reason expression in square brackets \eqref{L00_def} can be written as:
%\begin{eqnarray}
%&\int_{0}^{1}dt \dfrac{e^{-z_0^2/2}}{1-t^2}+\int_{1}^{\infty}dt\dfrac{e^{-z_0^2 t^2/2}}{1-t^2}+ \nonumber\\
%&+\dfrac{\pi}{2}e^{z_0^2/2}\erfi\dfrac{z_0}{\sqrt{2}}-\dfrac{\pi}{2}e^{z_0^2/2}\erfi\dfrac{z_0}{\sqrt{2}}
%\end{eqnarray}
Thus, the exact value $\mathcal{L}$ at point $(0,0)$ \eqref{L00_def} is:
\begin{equation}
\begin{split}
\alpha \mathcal{L}(0,0)=\lim\limits_{\epsilon\to +0}\left[\int_{0}^{1-\epsilon}dt \frac{1}{1-t^2}+\int_{1+\epsilon}^{\infty}dt\frac{1}{t\left(1-t^2\right)}\right]=
\\
=\lim\limits_{\epsilon\to +0}\frac{1}{2}\left(-\ln \epsilon+\ln 2+\ln\epsilon+\ln 2\right)=\ln 2
\end{split}
\end{equation}
On the other side the value at point $(0,0)$ can be obtained using approximation \eqref{C1}:
\begin{eqnarray}
&\alpha \mathcal{L}_{app}(0,0)=\alpha l(0)-\alpha \sqrt{\frac{2}{\pi \sigma^2}}\int_{-\infty}^{0}dz_0e^{\frac{-z_0^2}{2\sigma}}l(z_0)=\nonumber \\
&=A\left(1-\frac{\pi}{4}\right)+\frac{1}{2}\ln 2
\end{eqnarray}
Combing last results one can get the value of adjusting parameter 
\begin{eqnarray}
A=-\dfrac{\ln 2}{2-\frac{\pi}{2}}
\end{eqnarray}
\bibliography{AKhlyupin_TAslyamov_EffectiveSFPotential}
\end{document}